\documentclass[journal]{IEEEtran}
\usepackage{graphicx}
\usepackage{amsmath}
\usepackage{cite}
\usepackage{url}
\usepackage{booktabs}
\usepackage{tabularx}
\usepackage[table]{xcolor}
\usepackage{array}
\usepackage{makecell}
\usepackage{multirow}
\usepackage{algorithm}
\usepackage{algpseudocode}
\usepackage{float}

\graphicspath{{Figures/}}
\ifCLASSINFOpdf
\else
\fi
\begin{document}
%
\title{FDIFormer:Protocol-Aware Transformer Learning for False Data Injection Attack Detection in Smart Grid Networks   }
%
%
%

\author{Sandara Sathsarani Wijethunga,
        Muneeb Ul Hassan,~\IEEEmembership{Member,~IEEE,}
        and~Nasrin Sohrabi,~\IEEEmembership{Member,~IEEE}
\thanks{S. S. Wijethunga, M. U. Hassan, and N. Sohrabi are with the 
School of Information Technology, Deakin University, Australia 
(e-mail: s224740132@deakin.edu.au; muneeb.ulhassan@deakin.edu.au; nasrin.sohrabi@deakin.edu.au).}}

%
%

\markboth{Journal of \LaTeX\ Class Files,~Vol.~14, No.~8, August~2026}%
{Shell \MakeLowercase{\textit{et al.}}: Bare Demo of IEEEtran.cls for IEEE Journals}
%



\maketitle

\begin{abstract}
Smart grids use communication networks and intelligent electronic devices for reliable and automated operation of power systems. As these systems become increasingly interconnected, they are also becoming more exposed to cyberattacks such as message tampering, false command injection, and denial-of-service attacks. One particularly concerning threat is False Data Injection (FDI), where attackers manipulate communication messages by deleting, modifying, or adding packets. This is especially crucial in IEC 61850-based substations, where Generic Object-Oriented Substation Event (GOOSE) messages are used for delivering time-critical protection and control information between devices. Detecting FDI attacks in IEC 61850 GOOSE traffic remains challenging because malicious packets can closely resemble legitimate communication, making them difficult to distinguish from normal operational behaviour. Furthermore, many existing detection methods depend heavily on manually engineered protocol features, which require extensive domain knowledge and may not generalise well across different smart grid environments.
This paper proposes FDIFormer, a feature-engineering-free framework for FDI attack detection using structured textual representations of GOOSE packet sequences and fine-tuned pre-trained Transformer models. The proposed framework converts protocol packets into structured text windows that capture communication behaviour and enables Transformer models to learn attack-related patterns directly from the data.

The framework is evaluated using the QUT-ZSS-2023-GOOSE dataset under a scenario-level three-fold cross-validation strategy. Experimental results 
show that GraphCodeBERT achieves an MCC of 0.595 $\pm$ 0.122, achieving performance comparable to the strongest feature-engineered baseline, XGBoost (MCC = 0.604 $\pm$ 0.121), while improving MCC by 0.133 compared with the TF-IDF baselines. These findings demonstrate that pre-trained Transformer representations can provide an effective technique for FDI attack detection in IEC 61850 GOOSE communication without relying on manually engineered 
protocol features.

\end{abstract}


%
\IEEEpeerreviewmaketitle

\section{Introduction}

%
%
%
%
\IEEEPARstart{S}{mart} Grids have become an important part of modern power systems because they improve efficiency, reliability, and real-time monitoring of electricity networks \cite{hassan2019dpsurvey, liu2012cybersecurity}. With the switch of substations to digital architectures, standardised communication protocols have become essential for coordinating the growing number of interconnected devices \cite{iec61850}. The IEC 61850 standard has been widely used in modern substations to support communication between Intelligent Electronic Devices (IEDs), providing a common framework for interoperability and automation \cite{quincozes2024ereno, presekal2023attackgraph}.

The Generic Object-Oriented Substation Event (GOOSE) protocol is one of the communication services defined by IEC 61850, which is especially important for fast, event-driven exchange of protection and control information between substation devices \cite{iec61850}. GOOSE messages are developed for speed and reliable communication over shared Ethernet networks and are time-sensitive, not designed for security and thus subject to being hacked \cite{rajkumar2024cascading}.

IEC 61850 improves operational performance, but it also introduces cybersecurity challenges. Since GOOSE messages prioritise speed and reliability over security, attackers may exploit these weaknesses to manipulate communication traffic. One of the most critical threats is the False Data Injection (FDI) attack, where malicious data is inserted or modified to influence the operation of power system devices \cite{quincozes2024ereno, rajkumar2024cascading}. In such attacks, an adversary may delete, modify, or inject GOOSE messages to mislead protection relays or controllers into making incorrect decisions \cite{rajkumar2024cascading}. Successful FDI attacks can result in incorrect protection actions, equipment damage, or even power system outages.

To address these threats, researchers have proposed various intrusion detection systems based on machine learning and deep learning techniques \cite{li2022fdi}. Many existing approaches achieve strong detection performance using manually engineered protocol features. However, these methods often require significant domain expertise and can be difficult to adapt across different environments. In addition, the effectiveness of such systems is often highly dependent on the quality of the engineered features.

Recent advances in pre-trained language models have shown strong performance across a wide range of tasks beyond natural language processing \cite{min2023llmsurvey}. Models based on the Transformer architecture, including BERT \cite{devlin2019bert}, DistilBERT, CodeBERT \cite{feng2020codebert}, and GraphCodeBERT \cite{guo2021graphcodebert}, have shown a strong ability to learn useful patterns from structured sequential data. This suggests that GOOSE communication traffic can be represented as structured text and analysed using pre-trained Transformer models, reducing the need for manually designed features.

Using large language models for cybersecurity problems has attracted growing interest from researchers over recent years \cite{ferrag2024bertiot}. However, there is still limited research on fine-tuned pre-trained Transformer models for FDI detection in IEC 61850 GOOSE communication. Moreover, there is little evidence that clearly isolates how much of the detection performance is driven by pre-trained Transformer representations compared to baseline machine learning models using the same textual input representation.

To address this gap, this paper proposes FDIFormer, an end-to-end framework for detecting FDI attacks in IEC 61850 GOOSE traffic. FDIFormer transforms raw GOOSE packet sequences into structured textual representations and fine-tunes pre-trained Transformer models — including BERT \cite{devlin2019bert}, DistilBERT, CodeBERT \cite{feng2020codebert}, and GraphCodeBERT \cite{guo2021graphcodebert} — directly on these representations, eliminating the need for manual feature engineering. Unlike approaches that simply apply an existing model off the shelf, FDIFormer defines a complete pipeline that includes packet-to-text conversion, sequence windowing, and model fine-tuning, with GraphCodeBERT serving as the core encoder that captures both semantic and structural patterns in GOOSE traffic. FDIFormer is evaluated against engineered-feature baselines, TF-IDF baselines, and hybrid architectures under a common scenario-level cross-validation framework, and shows competitive detection performance while removing the dependency on domain-specific feature design \cite{wolf2020huggingface}. The contributions of this paper are as follows:

\begin{enumerate}

\item A structured textual representation is developed for IEC 61850 GOOSE packet sequences so that pre-trained Transformer models can process protocol traffic without any manual feature engineering.

\item FDIFormer is proposed as a complete detection pipeline that integrates packet-to-text conversion, sequence windowing, and fine-tuned Transformer-based classification, with GraphCodeBERT as the core encoder, enabling effective FDI attack detection directly from raw GOOSE traffic.

\item The proposed Transformer-based approach is compared with engineered-feature baselines and TF-IDF baselines using the same experimental framework.

\item The contribution of pre-trained Transformer representations is analysed by comparing Transformer models with classical machine learning models using the same structured text input.

\item Model performance is analysed across different FDI attack behaviours to understand which attack types are more difficult to detect.

\end{enumerate}

\section{Related Works}

This section reviews the studies that are most closely related to our work and highlights the key ideas that informed the development of the proposed approach. We start by discussing how large language models have been fine-tuned for specific domains, which forms the basis of our approach to adapt pre-trained Transformer models for analysing protocol traffic. Next we review the use of LLMs in cybersecurity and their growing use in smart grid applications. Finally, we study conventional smart grid intrusion detection methods based on manually engineered features, which are used as our baseline models. Together, these areas provide the background for this research and help identify the gap that FDIFormer aims to address.

\subsection{LLM Fine-Tuning for Domain-Specific Tasks}
Recent studies have proven that large language models can be adapted successfully to specialised areas through fine-tuning. Zhang et al. \cite{zhang2025cyberllama} enhanced a fine-tuned LLaMA-3.2-3B model with BiLSTM and CRF layers for cybersecurity named entity recognition, achieving an F1-score of 98.88\%. Their results demonstrate the potential of pre-trained language models when sufficient domain-specific training data is available.

Among the studies reviewed, the work of Saber et al. \cite{saber2026relay} s the most closely aligned with the objectives of this research. They converted relay measurements into structured text prompts and fine-tuned DistilBERT and GPT-2 for cyberattack detection in protective relays. Their results demonstrated that Transformer models can effectively learn from textual representations of power system data. Mehavilla et al. \cite{mehavilla2026llmids} compared fine-tuned LLMs with machine learning baselines using network flow data and reported that XGBoost still provided a strong balance between accuracy and efficiency. Yang et al. \cite{yang2025llmaptds} proposed LLM-APTDS for advanced persistent threat detection and demonstrated how LLMs can be used for complex cybersecurity reasoning tasks.

\subsection{LLMs for Cybersecurity}
Many researchers conducted research into the application of large language models in cybersecurity. Conceicao and Cruz \cite{conceicao2025maturity} tested several frontier LLMs on various types of cybersecurity tasks, demonstrating that the performance might greatly differ depending on the application domain. Yang et al. \cite{yang2025llmaemp}  introduced LLM-AE-MP which is based on DistilBERT embeddings, combined with autoencoder, LSTM and MLP for web attack detection. Zhang et al. \cite{zhang2024attackg} proposed AttacKG+ for converting threat intelligence reports into structured attack graphs, while Belcastro et al. \cite{belcastro2025klage} introduced KLAGE, which combines knowledge graphs with LLM-generated explanations to support explainable threat detection.

These studies show the increasing capability of LLMs in the cybersecurity field. However, their focus is mostly on general network security issues rather than packet-level FDI detection in IEC 61850 GOOSE communication.

\subsection{LLMs for Smart Grid Applications}

The use of large language models in smart grid applications remains at an early stage of development. Zaboli et al. \cite{zaboli2024llmgoose} evaluated anomaly detection in IEC 61850 communication using ChatGPT, Claude and Bard in a hardware-in-the-loop environment. Their study was interesting however mainly reliant on human interpretation without proper model training or comparison.

Other applications of LLMs are non-intrusive load monitoring \cite{chen2026nilm}, distributed energy resource monitoring \cite{zhao2025der}, semantic interoperability \cite{fatemi2025semantic}, and others in energy systems \cite{zhang2026energy}. These studies highlight growing interest in applying large language models to power system applications; however, none specifically investigates fine-tuned Transformer models for detecting FDI attacks in GOOSE traffic.

\subsection{Classical Smart Grid Security}
Most existing smart grid intrusion detection systems rely on manually engineered protocol features combined with traditional machine learning or deep learning models. Alsirhani et al. \cite{alsirhani2025ensemble} proposed an AI-based ensemble framework for smart grid intrusion detection, while Ijeh and Morsi \cite{ijeh2024bagging} applied fine tree bagging for attack classification. Ciaramella et al. \cite{ciaramella2025xai} used explainable deep learning for intrusion detection, and Zheng et al. \cite{zheng2025federated} proposed federated learning with unsupervised attack detection methods.

Although these methods show good performance, they generally depend on expert-designed protocol features and often require recalibration when applied to new environments. This creates challenges for scalability and deployment across different substations.

Overall, the study shows growing interest in both smart grid intrusion detection and Transformer-based learning. However, there is still limited research investigating fine-tuned pre-trained Transformer models for FDI attack detection in IEC 61850 GOOSE communication. Furthermore, the specific contribution of pre-trained Transformer representations has not been systematically evaluated against classical machine learning models using the same textual input representation.

\vspace{4pt}
\textit{Based on our review of the existing literature, we found no prior study that presents a complete end-to-end framework for converting raw IEC 61850 GOOSE packet sequences into structured textual representations and using fine-tuned pre-trained Transformer models to detect False Data Injection (FDI) attacks without manual feature engineering, as proposed by FDIFormer. In addition, existing studies have not systematically examined the contribution of pre-trained Transformer representations by comparing them with classical machine learning models using the same textual input. To the best of our knowledge, no previous work has also investigated detection performance across different FDI attack behaviours within a scenario-level cross-validation framework.}

\begin{table*}[!t]
\centering
\caption{Comparative Analysis of Existing Studies Related to Transformer Models, Smart Grid Cybersecurity, and False Data Injection Detection}
\label{tab:related_work_comparison}
\scriptsize
\renewcommand{\arraystretch}{1.0}
\setlength{\tabcolsep}{3pt}
\resizebox{\textwidth}{!}{
\begin{tabular}{|p{1.4cm}|p{0.55cm}|p{0.55cm}|p{2.2cm}|p{2.8cm}|p{2.4cm}|p{2.6cm}|p{2.8cm}|}
\hline
\textbf{Category} &
\textbf{Ref.} &
\textbf{Year} &
\textbf{Problem Focus} &
\textbf{Methodology} &
\textbf{Dataset / Platform} &
\textbf{Key Findings} &
\textbf{Research Gap} \\
\hline

\multirow{4}{=}{LLM Fine-Tuning}
& [1] & 2025
& Cybersecurity named entity recognition
& Fine-tuned LLaMA with BiLSTM and CRF layers
& Cybersecurity article corpus
& Achieved 98.88\% F1-score
& Focused on entity extraction, not intrusion detection in smart-grid traffic \\
\cline{2-8}

& [2] & 2025
& Advanced persistent threat detection
& LLM-based APT detection with graph reasoning
& Provenance and APT-related security data
& Improved attack detection precision and interpretability
& Not evaluated for IEC 61850 or smart-grid communication traffic \\
\cline{2-8}

& [3] & 2026
& Flow-based intrusion detection
& Fine-tuned LLMs compared with ML baselines
& Zeek network flow text data
& LLMs worked on structured network data, but XGBoost remained competitive
& Did not focus on IEC 61850 GOOSE protocol or FDI attacks \\
\cline{2-8}

& [4] & 2026
& Smart-grid relay cyberattack detection
& Relay measurements converted into structured prompts and fine-tuned using Transformer models
& Protective relay cyberattack dataset
& Textualised power-system data supported attack detection
& Does not analyse packet-level GOOSE traffic or FDI attack behaviour \\
\hline

\multirow{4}{=}{LLMs for Cybersecurity}
& [5] & 2025
& Cybersecurity task maturity analysis
& Benchmarking frontier LLMs across cybersecurity tasks
& Malware, honeypot, and CTF tasks
& Showed that LLM performance varies across cybersecurity tasks
& Not focused on smart-grid protocols or GOOSE traffic \\
\cline{2-8}

& [6] & 2025
& Web attack detection
& DistilBERT embeddings with autoencoder, LSTM, GAN, RL, and MLP
& Web attack datasets
& Improved web attack detection under imbalance
& Designed for web traffic, not industrial GOOSE communication \\
\cline{2-8}

& [7] & 2024
& Cyber threat intelligence extraction
& LLM pipeline for attack graph construction
& Cyber threat intelligence reports
& Converted unstructured reports into structured attack graphs
& Not a real-time packet-level intrusion detection system \\
\cline{2-8}

& [8] & 2025
& Explainable threat detection
& Knowledge graph, Graph-BERT, LIME, and LLM explanations
& Network security data
& Improved explainability for threat detection
& Does not address IEC 61850 GOOSE FDI detection \\
\hline

\multirow{5}{=}{LLMs in Smart Grids}
& [9] & 2024
& IEC 61850 anomaly analysis
& Evaluated ChatGPT, Claude, and Bard for smart-grid cybersecurity tasks
& Hardware-in-the-loop IEC 61850 test environment
& Showed early feasibility of LLMs for GOOSE-related analysis
& Exploratory study without systematic fine-tuning or model comparison \\
\cline{2-8}

& [10] & 2026
& Scene-aware energy load monitoring
& LLM-based non-intrusive load monitoring
& Smart-grid energy consumption data
& Improved contextual understanding of load behaviour
& Focused on energy monitoring, not cybersecurity or FDI detection \\
\cline{2-8}

& [11] & 2025
& Partial tripping detection in distributed energy resources
& LLM-based event interpretation and reasoning
& IEEE test network environment
& Improved event analysis for distributed energy resources
& Does not address intrusion detection in GOOSE communication \\
\cline{2-8}

& [12] & 2025
& Semantic interoperability in heterogeneous smart grids
& Fine-tuned LLaMA with LoRA and retrieval-augmented generation
& Smart-grid interoperability data
& Improved semantic integration between heterogeneous grid systems
& Does not focus on cyberattack detection or FDI analysis \\
\cline{2-8}

& [13] & 2026
& Review of LLM applications in energy systems
& Survey of LLM roles, enhancement methods, and application areas
& Energy systems literature
& Identified growing interest in LLMs for smart-grid applications
& Shows limited work on protocol-level smart-grid cybersecurity \\
\hline

\multirow{4}{=}{Smart Grid Security}
& [14] & 2025
& Intrusion detection in smart-grid environments
& AI-based ensemble intrusion detection framework
& Smart-grid intrusion detection datasets
& Achieved high detection accuracy using combined AI models
& Requires engineered features and labelled datasets \\
\cline{2-8}

& [15] & 2024
& Smart-grid cyberattack type classification
& Fine tree bagging ensemble with feature selection
& Physical and network feature dataset
& Achieved strong attack classification performance
& Depends on manually selected protocol and network features \\
\cline{2-8}

& [16] & 2025
& Explainable smart-grid intrusion detection
& CNN-based intrusion detection with explainable AI methods
& IEC protocol intrusion detection data
& Improved interpretability of deep learning-based detection
& Does not evaluate pre-trained Transformer representations \\
\cline{2-8}

& [17] & 2025
& Privacy-aware cyberattack detection
& Autoencoder-based learning with federated detection framework
& Distributed smart-grid data
& Improved privacy-preserving detection across distributed settings
& Not focused on pre-trained language models or GOOSE FDI detection \\
\hline

\end{tabular}
}
\end{table*}
\section{Preliminaries}
\subsection{IEC 61850 GOOSE Communication}

IEC 61850 is an international standard developed by the International Electrotechnical Commission (IEC) to support communication networks and systems used in power utility automation. It was introduced to overcome the limitations of specific substation communication protocols and to provide a common framework that allows Intelligent Electronic Devices (IEDs) from different suppliers to communicate and exchange information easily \cite{iec61850}. By defining standardised data models, communication services, and configuration mechanisms, IEC 61850 has become the foundation of modern digital substations, enabling higher levels of interoperability, automation, and operational reliability \cite{quincozes2024ereno}.

The standard consists of multiple parts covering areas such as system engineering, configuration, communication services, and protocol mappings. Among its communication services, Generic Object-Oriented Substation Event (GOOSE) and Sampled Values (SV) are the two most important for real-time operation. GOOSE is used to exchange time-critical protection and control information, while SV is used to transmit digitised measurements from instrument transformers. To meet the strict latency requirements of protection systems, both services operate directly over Ethernet multicast communication rather than relying on higher-layer protocols such as TCP/IP \cite{iec61850}.

GOOSE is particularly important because it enables the rapid exchange of protection and control signals between IEDs. It follows a publisher-subscriber communication model, where a publishing device broadcasts messages that can be received by multiple subscribing devices on the same network segment without establishing dedicated connections \cite{li2022fdi}. The GOOSE messages are retransmitted repeatedly after a state change and continue to be sent periodically during normal operation to improve reliability. This mechanism helps ensure that critical information is delivered even in the absence of acknowledgement messages \cite{quincozes2024ereno}.

Each GOOSE message is encoded using ASN.1 Basic Encoding Rules (BER) and contains a structured set of fields that represent the current state of the device. These fields play an important role in ensuring correct interpretation and synchronisation across communicating devices. The main fields include:

\begin{itemize}
    \item \textbf{APPID}: A 16-bit identifier used to uniquely distinguish each GOOSE publisher on the network.
    
    \item \textbf{stNum} (State Number): A counter that increases whenever the dataset changes, indicating a new state transition.
    
    \item \textbf{sqNum} (Sequence Number): A counter that increments with each retransmission within the same state, allowing receivers to track message order and freshness.
    
    \item \textbf{TimeAllowedToLive}: Defines the maximum time (in milliseconds) a subscriber should wait before considering the message lost, which may trigger a fail-safe action.
    
    \item \textbf{Dataset}: Contains the actual values being communicated, such as status information, quality indicators, and timestamps.
    
    \item \textbf{GOOSElength}: Represents the total size (in bytes) of the GOOSE protocol data unit (PDU).
\end{itemize}

The timing behaviour of GOOSE retransmissions can be described mathematically. Let $T_0$ represent the initial retransmission interval after a state change, and let $T_{max}$ be the maximum heartbeat interval. The retransmission interval $T_i$ at the $i$-th retransmission is given by:

\begin{equation}
T_i = \min(T_0 \cdot 2^i,\ T_{max})
\end{equation}

This exponential back-off mechanism ensures that messages are transmitted frequently immediately after a state change, when timely updates are most important, and less frequently during stable operating conditions to reduce network load. Any deviation from this expected pattern—such as irregular state transitions, unexpected sequence number changes, or abnormal timing behaviour—can be an indicator of potential cyberattacks \cite{quincozes2024ereno}.

Although GOOSE improves communication efficiency and reliability in substations, it was not originally designed with cybersecurity in mind. The messages are broadcast over Ethernet without built-in encryption or authentication, which makes them vulnerable to manipulation if an attacker gains access to the network. Although IEC 62351 later introduced security enhancements for IEC 61850 systems, its adoption in practice remains limited due to additional processing overhead and the strict timing constraints of protection systems \cite{iec62351}. As a result, operational GOOSE traffic in many substations is still exposed to potential network-level attacks within the local area network.

\subsection{False Data Injection Attacks}

False Data Injection (FDI) attacks are considered as one of the most serious cyber threats for industrial communication systems, including IEC 61850-based substations \cite{quincozes2024ereno}. FDI attacks are more subtle than denial-of-service attacks, which aim to stop communication entirely. They operate by changing the content or flow of valid messages thus monitoring systems and protection relays perform faulty operational decisions, while appearing as normal network activity \cite{li2022fdi, presekal2023attackgraph}.

The attacks on FDI in IEC 61850 GOOSE communication are generally categorised into three major types based on the attacker's manipulation of the messages:

\subsubsection{Message Injection (Addition)}

In this type of attack, the adversary pretends to be a legitimate IED publisher to produce and inject fake GOOSE packets into the network. In these forged messages, invalid values may be inserted for the datasets. For instance, a false breaker trip command, which can cause unintended actions in subscribing devices. Let $m_t$ be a valid GOOSE message at time $t$ in formal terms. An injection attack produces a false message $\tilde{m}_t$ such that:

\begin{equation} 
\tilde{m}_t = m_t + \delta_t 
\end{equation}

where $\delta_t$ denotes the malicious changes introduced by the attacker to the dataset or protocol fields.

\subsubsection{Message Modification} 
 
In a modification attack, the attacker intercepts a valid GOOSE message and changes one or more of its fields before sending the message to its intended recipients. 

These changes can impact the values within a dataset, sequence data such as stNum or sqNum, or even timing fields. This process can be summarised as:

\begin{equation} 
m'_t = f(m_t) 
\end{equation} 

where $f(\cdot)$ denotes the malicious transformation of the original message $m_t$. The modified packet would still be valid with respect to the correct GOOSE structure and often hard to detect with traditional signature-based techniques. 

\subsubsection{Message Deletion (Suppression)}

For a deletion or suppression attack, the adversary selectively drops GOOSE messages before they reach their intended subscribers. This can be especially harmful if messages that correspond to a real state change are missing, as this can cause devices to assume a fault or to initiate fail-safe behaviour. Let $\mathcal{M}$ be the set of expected GOOSE messages in time window $[t_1, t_2]$ A deletion attack leads to a received subset $\mathcal{M}' \subset \mathcal{M}$ such that:

\begin{equation} |
\mathcal{M'}| < |\mathcal{M}| 
\end{equation}

which implies that the attacker has deleted some messages.

Successful FDI attacks may have an adverse effect on IEC 61850 GOOSE communication. They may lead to unnecessary operations of breakers, malfunctioning of protection devices, damage to equipment and even cascading failures throughout the power system \cite{presekal2023attackgraph, rajkumar2024cascading}. Furthermore, these attacks are difficult to detect because FDI traffic generally preserves the same structural format as legitimate GOOSE messages, making it difficult to distinguish malicious activity from normal operational communication \cite{quincozes2024ereno}.

\subsection{Pre-Trained Transformer Models}

Transformer models, first introduced by Vaswani et al. \cite{vaswani2017attention}, are a family of deep learning architectures based on self-attention mechanisms that enable the learning of relationships between elements in a sequence regardless of their distance from  each other. Unlike recurrent models such as LSTMs and GRUs, Transformers do not process tokens in a sequence sequentially, but in parallel. This enables more efficient training on large datasets and improves their ability to capture long-range dependencies \cite{min2023llmsurvey}.

At the core of this architecture is the self-attention mechanism. This mechanism receives a sequence of tokens and returns a weighted representation of each token in the sequence relative to all the other tokens. The attention operation takes as input a query matrix $\mathbf{Q}$, a key matrix $\mathbf{K}$ and a value matrix $\mathbf{V}$. It is computed as follows:

\begin{equation} 
\text{Attention}(\mathbf{Q}, \mathbf{K}, 
\mathbf{V}) = \text{softmax}\!\left( \frac{\mathbf{Q}\mathbf{K}^\top} {\sqrt{d_k}}\right)\mathbf{V} 
\end{equation} 

Here, $d_k$ is the dimension of the key vectors and the scaling factor $\frac{1}{\sqrt{d_k}}$ is used to prevent very large dot-products, which make training unstable \cite{vaswani2017attention}. In practice, multiple attention heads are used in parallel so that the model can attend to different aspects of the input simultaneously:

\begin{equation} 
\text{MultiHead}(\mathbf{Q}, \mathbf{K}, 
\mathbf{V}) = \text{Concat}(\text{head}_1, 
\ldots, \text{head}_h)\mathbf{W}^O, 
\end{equation} 

where each head$_i$ is defined as Attention$(\mathbf{Q}\mathbf{W}_i^Q,\, \mathbf{K}\mathbf{W}_i^K,\, \mathbf{V}\mathbf{W}_i^V)$, and $\mathbf{W}_i^Q$, $\mathbf{W}_i^K$, $\mathbf{W}_i^V$, and $\mathbf{W}^O$ are learnable parameters.

Pre-trained Transformer models build on this architecture by pre-training on large-scale datasets with self-supervised learning objectives and then fine-tuning the learned representations on specific downstream tasks \cite{wolf2020huggingface}. This two-stage process allows the models to learn general-purpose representations during pre-training and then to adapt efficiently to new tasks with relatively small labelled datasets. The models used in this work are summarised below.

\subsubsection{BERT} BERT (Bidirectional Encoder Representations from Transformers) \cite{devlin2019bert} proposed bidirectional pre-training for language models. Rather than processing text in one direction, BERT learns context from both sides of a token (left and right). It is trained using Masked Language Modelling (MLM), where some tokens are masked and the model predicts them, and Next Sentence Prediction (NSP), where the model learns whether two sentences follow each other in a text. This bidirectional design allows BERT to produce rich contextual embeddings which are useful in many NLP tasks. 

\subsubsection{DistilBERT} 
DistilBERT \cite{sanh2019distilbert} is a lighter and faster BERT version obtained by knowledge distillation. It is ~40\% smaller and ~60\% faster than BERT-base, however maintains most of its performance. The model removes some components like token-type embeddings and the pooler layer, and reduces the number of transformer layers from 12 to 6. This makes it more feasible when computational resources are limited. 

\subsubsection{RoBERTa} 
RoBERTa (Robustly Optimised BERT Pretraining Approach) \cite{liu2019roberta} is an improvement over BERT by optimising the training process. It removes the NSP objective, uses larger batch sizes, is trained on more data, and uses a larger byte-pair encoding vocabulary. These changes enable the model to better use its capacity and consistently improve its performance on benchmarks. 

\subsubsection{ELECTRA} 
ELECTRA \cite{clark2020electra} proposes a more efficient pre-training method called Replaced Token Detection. Instead of masking tokens , the small generator replaces some tokens and a discriminator is trained to identify which tokens were replaced . The model is trained on all tokens rather than masked ones, thus achieving strong performance with much improved sample efficiency.

\subsubsection{CodeBERT} 
CodeBERT \cite{feng2020codebert} is a bimodal pre-trained model trained on both source code and natural language descriptions. During pre-training, it uses objectives such as MLM and replaced token detection. Its strength in modelling structured and syntax-rich inputs makes it an ideal solution to represent GOOSE packet sequences, which are also expressed in a structured key-value format similar to code.

\subsubsection{GraphCodeBERT}
GraphCodeBERT \cite{guo2021graphcodebert} improves CodeBERT by adding structural information from data flow graphs. It also learns the information flow between variables in a sequence, in addition to token-level relations. This enables the model to learn deeper structural dependencies. It is particularly suitable for analysing GOOSE packet representations because it can model structured relationships. 

\subsubsection{ModernBERT}
ModernBERT is a recent encoder-only Transformer architecture with several architectural improvements, including rotary positional embeddings, a combination of local and global attention layers, and an extended context window of up to 8,192 tokens. These improvements make it more suitable for longer structured sequences such as multi-packet GOOSE windows.

In this work, all seven Transformer models described above are fine-tuned using structured textual representations of GOOSE packet sequences. The implementation uses the HuggingFace Transformers framework \cite{wolf2020huggingface}. This configuration allows us to evaluate the performance of these models in FDI attack detection, without relying on manual feature engineering in IEC 61850 GOOSE communication.

\subsection{Matthews Correlation Coefficient (MCC)}

Model performance is evaluated using the Matthews Correlation Coefficient (MCC). Unlike accuracy, which can be misleadingly high when a model predicts only the majority class. MCC incorporates all elements of the confusion matrix and produces values ranging between $-1$ 
and $+1$. Given the number of true positives (TP), true negatives (TN), false positives (FP), and false negatives (FN), MCC is defined as \cite{chicco2021mcc}:

\begin{equation}
\footnotesize
\text{MCC} = \frac{TP \times TN - FP \times FN}
{\sqrt{(TP+FP)(TP+FN)(TN+FP)(TN+FN)}}
\end{equation}
A value of 0 represents random performance, and $+1$ represents perfect prediction. Therefore, MCC is more suitable  of realistic detection capability in imbalanced binary classification cases.

\section{FDIFormer System Model}

This section describes the overall design and architecture of the proposed FDIFormer framework. As illustrated in Fig.~\ref{fig:system_model}, the FDIFormer framework consists of three main components: the Smart Grid Substation Environment, the FDI Attack Layer, and the Transformer-based Detection System, supported by an offline training pipeline. 

\subsection{Smart Grid Substation Environment}

The considered environment is a digital substation where multiple Intelligent Electronic Devices (IEDs) communicate continuously over a substation local area network (LAN) using IEC 61850 GOOSE messages \cite{quincozes2024ereno}. As shown in Zone 1 of Fig.~\ref{fig:system_model}, the setup includes three IEDs: two transformer protection units (XFMR1 and XFMR2) and one feeder protection unit (FDR). These devices are connected through an Ethernet switch and exchange protection and control information using the GOOSE publisher-subscriber mechanism \cite{iec61850}.

During normal operation, each IED broadcasts GOOSE messages periodically based on a configured heartbeat interval and also transmits rapid updates whenever a state change occurs. All traffic passing through the Ethernet switch is passively captured using a network tap and packet sniffer deployed on the substation LAN. The captured packets are stored as PCAP files and later processed by the FDIFormer pipeline. Since the monitoring is fully passive, the system does not introduce any additional delay into the protection communication path, making it suitable for strict real-time substation requirements \cite{li2022fdi}.

\subsection{Adversary Model}

The threat model is illustrated as Zone 2 in Fig.~\ref{fig:system_model}. We assume that the attacker has gained access to the substation local area network and can monitor, intercept, and manipulate GOOSE traffic as it passes through the Ethernet switch. The attacker does not have physical access to any IED and cannot directly control or reconfigure these devices.

As mentioned before, the attacker can perform three main types of FDI attacks: message injection, message modification, and message deletion. The goal of the attacker is to influence the behaviour of protection relays and IEDs, potentially leading to incorrect breaker operations, suppression of valid trip signals, equipment damage, or even cascading failures across the power system \cite{presekal2023attackgraph, rajkumar2024cascading}.

A common characteristic of these attacks is that they are designed to remain stealthy. The manipulated packets still follow the same structural format as legitimate GOOSE messages, which makes them difficult to detect using only rule-based or protocol conformance checks. As a result, detecting such attacks requires learning-based methods that can capture subtle patterns across sequences of packets rather than relying only on individual message inspection \cite{presekal2023attackgraph}.

\subsection{Motivation}

Most existing intrusion detection systems for IEC 61850 GOOSE traffic rely on manually engineered protocol features. These approaches often require deep domain knowledge, are time-consuming to design, and do not generalise well across different substations \cite{presekal2023attackgraph}. In addition, the effectiveness of these approaches depends heavily on the quality of manually engineered features, meaning that previously unseen attack patterns may remain undetected if they are not represented by those features\cite{li2022fdi}.

The main motivation behind FDIFormer is to remove this dependency on manual feature engineering. Instead, raw GOOSE packet sequences are converted into structured textual representations and analysed using pre-trained Transformer models. As discussed earlier, the structured key-value format of GOOSE traffic is similar to the syntax of programming languages, which makes it well-suited for code-aware models such as GraphCodeBERT \cite{guo2021graphcodebert}. This provides a natural opportunity to apply pre-trained representations for FDI detection without relying on domain-specific feature design.

\subsection{System Model and Structure}

As shown in Fig.~\ref{fig:system_model}, the FDIFormer framework consists of three main components: (1) the Smart Grid Substation Environment, (2) the FDI Attack Layer, and (3) the Transformer-based Detection System, supported by an offline training pipeline. The data flow between these components is described below.

\subsubsection{Packet Capture Module}
GOOSE traffic is captured passively from the substation network using a network tap and packet sniffer. The captured PCAP files are forwarded to the packet processing engine, where relevant IEC 61850 fields are extracted from each packet. These include APPID, stNum, sqNum, GOOSElength, num\_of\_data, and payload information. Additional derived features such as $\Delta$stNum, $\Delta$sqNum, state and sequence reset flags, and inter-packet time\_delta are also computed. APPID values are mapped to their corresponding IED names (XFMR1, XFMR2, FDR) to improve interpretability and generalisation.

\subsubsection{Numerical Feature Set}
The packet processing engine generates a numerical feature vector $F_w$ for each packet window, as shown in Fig.~\ref{fig:system_model}. This includes state and sequence variation features, reset indicators (st\_reset, sq\_reset), length-based anomalies, payload change indicators, and other aggregated statistics. The feature set $F_w$ is used by traditional machine learning baselines and hybrid models in the experimental evaluation.

\subsubsection{Structured Text Generator}
Each packet window is transformed into a structured text representation by the structured text generator. Five consecutive packets from the same IED are grouped using a sliding window with stride three, forming a window-level representation $T_w$. Each window is prefixed with the IED identifier and an estimated attack-type label (Normal, Structural, Value, or Sequence), which is derived from observed anomaly patterns within the window. This prefix provides additional contextual information to the Transformer model before processing individual packet fields.

\subsubsection{Detection Engine}
The structured text representation $T_w$ is tokenised and passed into a fine-tuned Transformer model. As shown in Fig.~\ref{fig:system_model}, the framework supports several pre-trained models, including BERT, RoBERTa, DistilBERT, CodeBERT, GraphCodeBERT, ELECTRA, and ModernBERT. Each model is fine-tuned as a binary classifier for detecting FDI attacks. The decision module applies a probability threshold $\theta$, selected using validation performance based on MCC, to produce the final prediction $\hat{y} \in \{0,1\}$, where 0 represents benign traffic and 1 represents an attack.

\subsubsection{Alert and Monitoring Module}
If a window is classified as an attack, the alert module generates logs and notifications for operators. The system outputs include predicted labels (benign or FDI), attack category information, event logs, alerts, and performance summaries for monitoring and analysis.

\subsection{Offline Training Environment}

\begin{figure*}[!t]
    \centering
    \includegraphics[width=\textwidth]{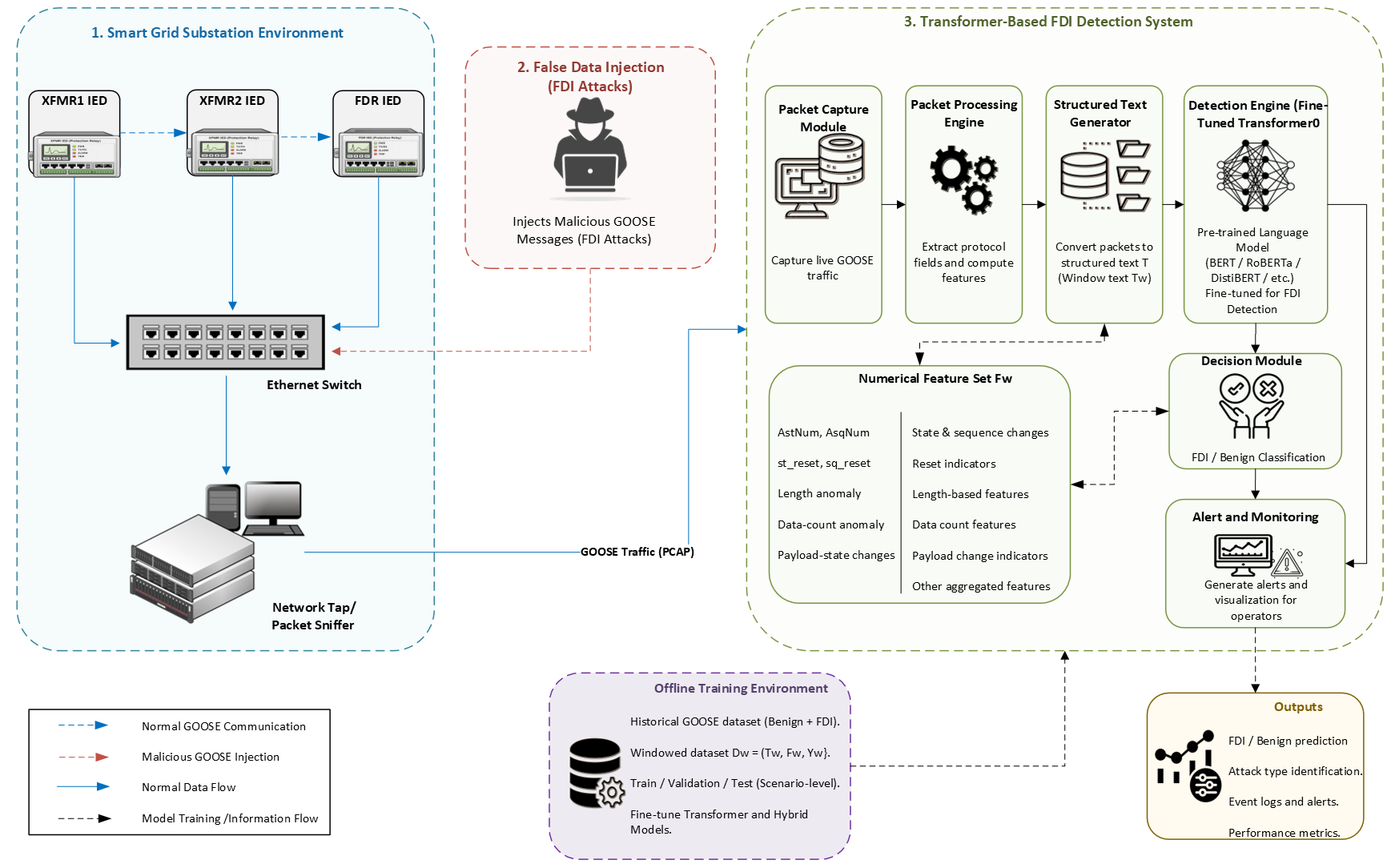}
    \caption{Overview of the proposed FDIFormer framework for detecting False Data Injection attacks in IEC 61850 GOOSE traffic. The framework transforms GOOSE packet sequences into structured text windows and evaluates Transformer, engineered-feature, TF-IDF, and hybrid models for attack classification.}
    \label{fig:system_model}
\end{figure*}

As shown at the bottom of Fig.~\ref{fig:system_model}, the FDIFormer model is trained offline using historical GOOSE datasets containing both normal and attack traffic. The dataset is first converted into windowed samples $D_w = \{T_w, F_w, Y_w\}$ and then split into training, validation, and test sets at the scenario level to avoid data leakage.

The pre-trained Transformer models and hybrid baselines are trained using this offline dataset and later deployed for inference on unseen or live traffic. In this work, the QUT-ZSS-2023-GOOSE dataset \cite{qut2023dataset} is used for all training and evaluation experiments.

Fig.~\ref{fig:system_model} illustrates the overall FDIFormer:Protocol-Aware Transformer Learning system model used throughout this research.

\section{Methodology of FDIFormer}
This research follows a quantitative experimental methodology to evaluate the effectiveness of pre-trained Transformer models for detecting False Data Injection (FDI) attacks in IEC 61850 GOOSE communication. The overall framework consists of data preparation, structured text generation, model training, and performance evaluation.

Fig.~\ref{fig:methodology_flowchart} shows the overall methodology used in this study.

\subsection{Methodology Overview}
\begin{figure}[!t]
    \centering
    \includegraphics[width=1.00\columnwidth]{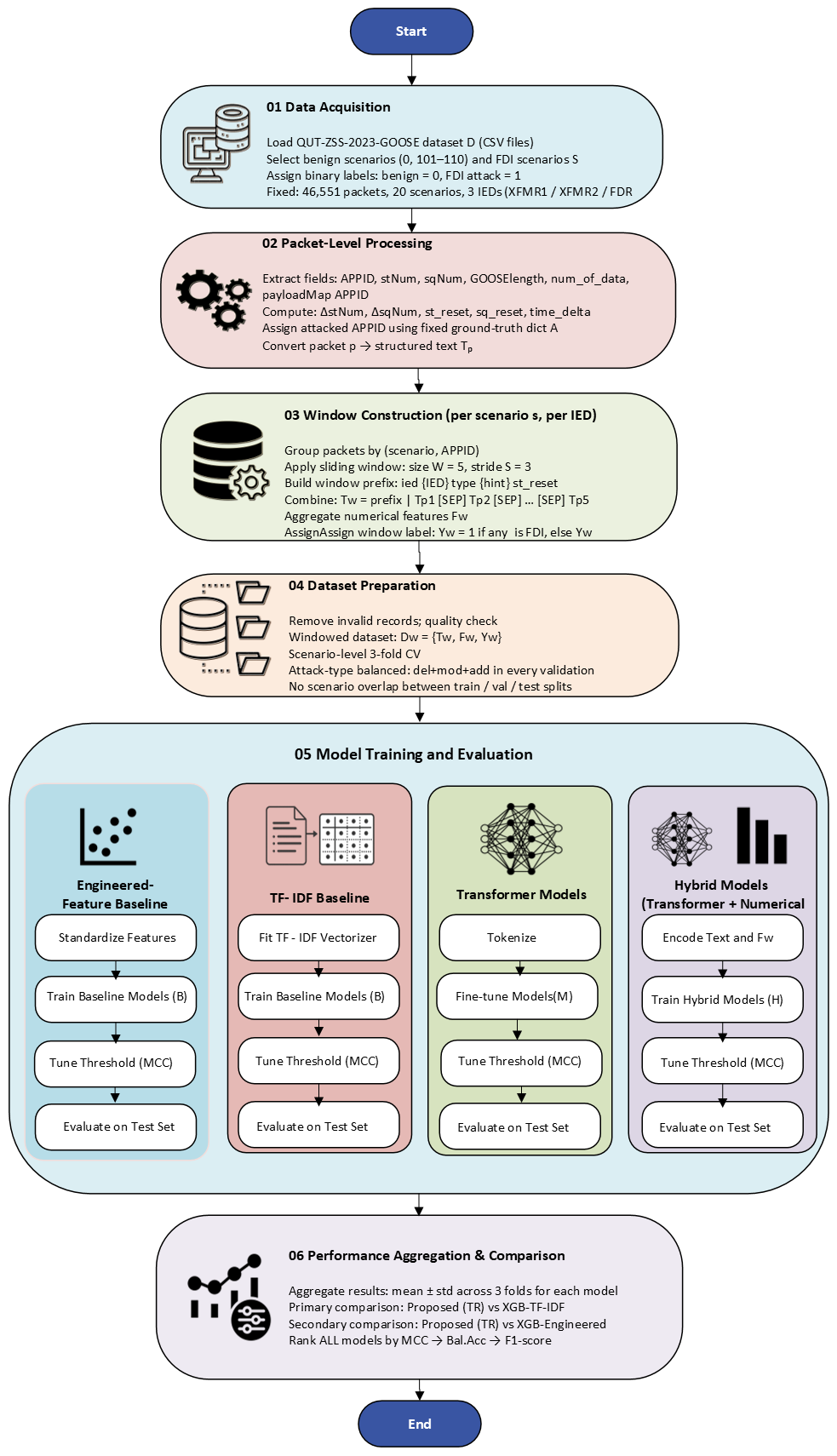}
    \caption{Methodology flowchart of the proposed Transformer-based FDI detection framework.}
    \label{fig:methodology_flowchart}
\end{figure}

The proposed FDIFormer pipeline consists of six main stages, as illustrated in Figure~\ref{fig:methodology_flowchart}.

\textbf{Stage 1 — Data Acquisition:}  
The QUT-ZSS-2023-GOOSE dataset \cite{qut2023dataset} is loaded from CSV files. It includes 11 benign operational scenarios and 9 FDI attack scenarios, identified in the raw dataset using scenario codes (e.g., 821-823, 841-843, and 861-863), as detailed in Table~II for the attack scenarios). Each record is labelled in a binary format, where benign traffic is assigned label 0 and FDI attacks are assigned label 1. In total, the dataset contains 46,551 packets collected from three IEDs (XFMR1, XFMR2, and FDR) across 20 scenarios.

\textbf{Stage 2 — Packet Processing:}  
Each GOOSE packet is parsed to extract key protocol fields such as APPID, stNum, sqNum, GOOSElength, num\_of\_data, and payload values. Additional derived features are also computed, including changes in state and sequence numbers, reset indicators, and time differences between packets. The APPID is mapped to a meaningful IED name, and each packet is then converted into a structured text format.

\textbf{Stage 3 — Window Construction:}  
Packets are grouped by IED and scenario, and a sliding window approach is applied with a window size of 5 and stride of 3. This produces both structured text windows ($T_w$) and numerical feature sets ($F_w$). Each window is also prefixed with the corresponding IED name and a high-level indication of the expected behaviour (e.g., normal or attack-related). A window is labelled as an attack if at least one packet inside it contains FDI activity.

\textbf{Stage 4 — Data Preparation:}  
After windowing, the dataset is cleaned by removing invalid or overlapping records. The final dataset $D_w = \{T_w, F_w, Y_w\}$ is then split using a scenario-level three-fold cross-validation strategy. This ensures that no scenario appears in more than one of training, validation, or testing sets, preventing data leakage.

\textbf{Stage 5 — Model Training and Evaluation:}  
Four groups of models are evaluated under the same conditions: engineered-feature models using $F_w$, TF-IDF models using $T_w$, Transformer-based models using fine-tuned pre-trained encoders, and hybrid models combining both text and numerical features. For each model, the decision threshold is tuned using the validation set to maximise MCC, and final evaluation is performed on unseen test data.

\textbf{Stage 6 — Results Comparison:}  
Results are reported as the mean and standard deviation across all folds to provide both average performance and variability. The main comparison focuses on Transformer models versus TF-IDF baselines using the same text input, which highlights the benefit of pre-trained representations. Additional comparisons are made against engineered-feature models to evaluate performance differences across feature types.

\subsection{Structured Text Representation}
A key part of this work is converting raw GOOSE packets into structured text, which allows Transformer models to process network traffic without manual feature engineering.

Each packet is transformed into a sequence containing:

\begin{itemize}
    \item The IED name derived from APPID (XFMR1, XFMR2, FDR)
    \item State and sequence numbers (stNum, sqNum)
    \item Changes in state and sequence values
    \item Reset indicators for state and sequence
    \item Protocol length and dataset size information
    \item Payload variation indicators
    \item Binary anomaly flags based on training statistics
    \item Time difference between packets
\end{itemize}

Five consecutive packets are then combined using a sliding window to form $T_w$. This window-based representation enables the model to capture both short-term behaviours and longer sequential patterns within network traffic.

The label for each window is defined as:

\begin{equation}
Y_w =
\begin{cases}
1 & \text{if any packet in the window is FDI} \\
0 & \text{otherwise}
\end{cases}
\end{equation}

\subsection{Cross-Validation Strategy}
A scenario-level three-fold cross-validation approach was used to ensure a fair evaluation Training, validation, and testing scenarios were separated at the scenario level to prevent data leakage between folds. All anomaly statistics and threshold selection procedures were calculated using training data only. Table II details the fold assignments.

\begin{table}[H]
\centering
\caption{Scenario-level distribution of FDI attack types across the three cross-validation folds.}
\label{tab:fold_distribution}
\scriptsize
\renewcommand{\arraystretch}{1.2}
\setlength{\tabcolsep}{2pt}

\resizebox{\columnwidth}{!}{
\begin{tabular}{|c|p{3.1cm}|p{2.4cm}|p{2.4cm}|}
\hline
\textbf{Fold} &
\textbf{Training FDI Scenarios} &
\textbf{Validation FDI Scenarios} &
\textbf{Testing FDI Scenarios} \\
\hline

1 &
842 (Modification), 843 (Addition), 861 (Deletion), 862 (Modification) &
821 (Deletion), 822 (Modification), 823 (Addition) &
841 (Deletion), 863 (Addition) \\
\hline

2 &
821 (Deletion), 841 (Deletion), 843 (Addition), 862 (Modification) &
823 (Addition), 842 (Modification), 861 (Deletion) &
822 (Modification), 863 (Addition) \\
\hline

3 &
841 (Deletion), 861 (Deletion), 862 (Modification), 863 (Addition) &
821 (Deletion), 842 (Modification), 843 (Addition) &
822 (Modification), 823 (Addition) \\
\hline

\end{tabular}
}
\end{table}

\subsection{Model Development}

To evaluate the effectiveness of different learning approaches for FDI attack detection, three model categories were investigated: Transformer-based models, classical machine learning baselines, and hybrid Transformer-numerical architectures.

\subsubsection{Transformer-Based Models}

The first model category consists of fine-tuned pre-trained Transformer models. Structured text windows generated from GOOSE packet sequences were used as inputs to BERT-base, DistilBERT, RoBERTa, ELECTRA-small, CodeBERT, GraphCodeBERT, and ModernBERT. Each model was fine-tuned as a binary classifier to detect between benign and FDI attack traffic.

\begin{algorithm}[!t]
\caption{Proposed Transformer-Based Detection of FDI Attacks in GOOSE Traffic}
\label{alg:transformer}
\begin{algorithmic}[1]

\Require Dataset $D$, scenario set $S$, Transformer model set $M_T$, window size $W$, stride $r$, folds $K$

\Ensure Transformer predictions $\hat{Y}_T$ and evaluation results $R_T$

\State Load GOOSE traffic files from dataset $D$
\State Select benign and FDI attack scenarios from $S$
\State Assign binary labels: benign $\rightarrow 0$, FDI attack $\rightarrow 1$

\For{each GOOSE packet $p \in D$}
    \State Extract APPID, stNum, sqNum, GOOSElength, num\_of\_data, and payload values
    \State Compute $\Delta$stNum, $\Delta$sqNum, st\_reset, sq\_reset, length anomaly, payload anomaly, and time\_delta
    \State Map APPID to IED semantic name
    \State Construct attack-type hint
    \State Convert packet $p$ into structured text representation $T_p$
\EndFor

\For{each scenario $s \in S$}
    \State Group packets by scenario and APPID
    \State Apply sliding window of size $W$ with stride $r$
    \State Construct window-level text representation $T_w$
    \State Assign label $Y_w = 1$ if any packet in the window is FDI; otherwise $Y_w = 0$
\EndFor

\State Split text-window dataset using scenario-level $K$-fold cross-validation

\For{each fold $k = 1$ to $K$}
    \State Separate scenarios into train, validation, and test sets
    \State Compute class weights using training-fold counts only

    \For{each Transformer model $m \in M_T$}
        \State Tokenise $T_w$ using the tokenizer of model $m$
        \State Fine-tune $m$ for binary classification
        \State Tune threshold $\theta$ on validation set using MCC
        \State Predict on unseen test scenarios
        \State Compute MCC, Balanced Accuracy, F1-score, ROC-AUC, and PR-AUC
    \EndFor

\EndFor

\State Aggregate results across all folds

\Return Final Transformer evaluation results $R_T$

\end{algorithmic}
\end{algorithm}

\subsubsection{Classical Machine Learning Baselines}

The classical machine learning baselines evaluated two baseline approaches. The first used manually engineered protocol features extracted from GOOSE traffic and trained using XGBoost and LightGBM classifiers. The second used TF-IDF representations generated from the same structured text windows and applied the same machine learning algorithms. These baselines were included to evaluate the contribution of pre-trained Transformer representations.

\begin{algorithm}[!t]
\caption{Classical Machine Learning Baseline Evaluation}
\label{alg:baseline}
\begin{algorithmic}[1]
\Require Windowed dataset $D_w$, engineered features $F_w$, text windows $T_w$, baseline models $B$, folds $K$
\Ensure Baseline evaluation results $R_B$

\State Split $D_w$ using scenario-level $K$-fold cross-validation

\For{each fold $k=1$ to $K$}
    \State Separate training, validation, and testing scenarios
    \State Fit all preprocessing steps using training data only
    \State Fit TF-IDF vectoriser using training text windows only

    \For{each baseline model $b \in B$}
        \State Train $b$ using engineered numerical features $F_w$
        \State Tune threshold $\theta$ on validation data using MCC
        \State Evaluate $b$ on unseen test scenarios

        \State Train $b$ using TF-IDF text features
        \State Tune threshold $\theta$ on validation data using MCC
        \State Evaluate $b$ on unseen test scenarios

        \State Record MCC, Balanced Accuracy, F1-score, ROC-AUC, and PR-AUC
    \EndFor
\EndFor

\State Aggregate fold-wise results as mean $\pm$ standard deviation
\Return Final baseline results $R_B$
\end{algorithmic}
\end{algorithm}

\subsubsection{Hybrid Transformer-Numerical Models}

The hybrid architectures combine Transformer representations with engineered numerical protocol features. Transformer embeddings generated from the text windows were concatenated with numerical features before classification. 

\begin{algorithm}[!t]
\caption{Hybrid Transformer-Numerical FDI Detection}
\label{alg:hybrid}
\begin{algorithmic}[1]
\Require Windowed dataset $D_w=\{T_w,F_w,Y_w\}$, hybrid model set $M_H$, folds $K$
\Ensure Hybrid evaluation results $R_H$

\State Split $D_w$ using scenario-level $K$-fold cross-validation

\For{each fold $k=1$ to $K$}
    \State Separate training, validation, and testing scenarios
    \State Standardise numerical features using training data only
    \State Compute class weights using training labels only

    \For{each hybrid model $h \in M_H$}
        \State Tokenise structured text windows $T_w$
        \State Encode $T_w$ using the Transformer branch
        \State Process numerical features $F_w$ using numerical branch
        \State Concatenate Transformer embedding and numerical representation
        \State Train hybrid classifier using weighted loss
        \State Predict attack probabilities on validation data
        \State Tune decision threshold $\theta$ using validation MCC
        \State Predict labels on unseen test scenarios
        \State Compute MCC, Balanced Accuracy, F1-score, ROC-AUC, and PR-AUC
    \EndFor
\EndFor

\State Aggregate fold-wise results as mean $\pm$ standard deviation
\Return Final hybrid results $R_H$
\end{algorithmic}
\end{algorithm}

This approach was evaluated to determine whether combining learned textual representations and protocol-specific numerical information could improve detection performance.

\subsection{Model Training Configuration}

Seven pre-trained Transformer architectures are fine-tuned: BERT-base-uncased, DistilBERT-base-uncased, RoBERTa-base, ELECTRA-small-discriminator, GraphCodeBERT-base, CodeBERT-base, and ModernBERT-base. All models share identical hyperparameters: learning rate $2 \times 10^{-5}$, cosine annealing scheduler with 10\% warmup, maximum 7 training epochs, early stopping patience 3 on validation MCC, label smoothing 0.1, weighted cross-entropy loss proportional to inverse class frequency, and maximum input sequence length 512 tokens. All experiments are implemented in Python using the HuggingFace Transformers library \cite{wolf2020huggingface}, PyTorch, XGBoost, LightGBM, and Scikit-learn, and are executed on Google Colab with an NVIDIA A100 GPU.

\subsection{Performance Evaluation}
Model performance was evaluated using Matthews Correlation Coefficient (MCC), Balanced Accuracy, F1-score, ROC-AUC, and PR-AUC. MCC was selected as the primary evaluation metric because it remains reliable under class imbalance and provides a balanced assessment of classification performance.

The final results were obtained by averaging the performance across all cross-validation folds. The evaluation focuses on comparing Transformer models with engineered-feature baselines, TF-IDF baselines, and hybrid architectures while also analysing behaviour across different FDI attack types.
\section{Experimental Evaluation}

\subsection{Experimental Environment}
All experiments were conducted in Google Colab using an NVIDIA A100 40GB GPU. Transformer models are implemented using HuggingFace Transformers, PyTorch, and the datasets library. Classical baselines use Scikit-learn, XGBoost, and LightGBM. All code is version-controlled using Git/GitHub, with fixed random seeds (seed = 42) to ensure full reproducibility. Tokenisation uses each model’s associated HuggingFace tokeniser with padding and truncation to 512 tokens. The QUT-ZSS-2023-GOOSE dataset [18] is processed using pandas and NumPy, with structured text windows generated via a custom Python pipeline.
\subsection{Experiment Setup}
We evaluate 15 models organised into four categories: (i) seven fine-tuned Transformer models; (ii) four hybrid Transformer-numerical models; (iii) two expert-engineered ML baselines (XGBoost and LightGBM with 36 hand-crafted features); and (iv) two same-text TF-IDF baselines (XGBoost and LightGBM on identical text windows). All models are evaluated under the same scenario-level three-fold cross-validation framework described in Section V-C, with the same data splits, class weighting, and threshold optimisation procedure. Performance is reported as mean ± standard deviation across three folds for MCC, Balanced Accuracy, and F1-score, and as mean ROC-AUC.
\subsection{Overall Performance}
Table III reports the complete performance of all 15 models ranked by mean MCC. FDIFormer with GraphCodeBERT achieves MCC = 0.595 ± 0.122, which is statistically equivalent to XGBoost-Engineered (0.604 ± 0.121; difference 0.009, within the standard deviation of both models), while requiring zero protocol-specific feature engineering. TF-IDF baselines are clearly separated below all other model families. Fig. 3 visualises the full MCC ranking with error bars; Fig. 4 presents the performance fingerprint heatmap across all four metrics.
\begin{figure}[H]
    \centering
    \includegraphics[width=\columnwidth]
    {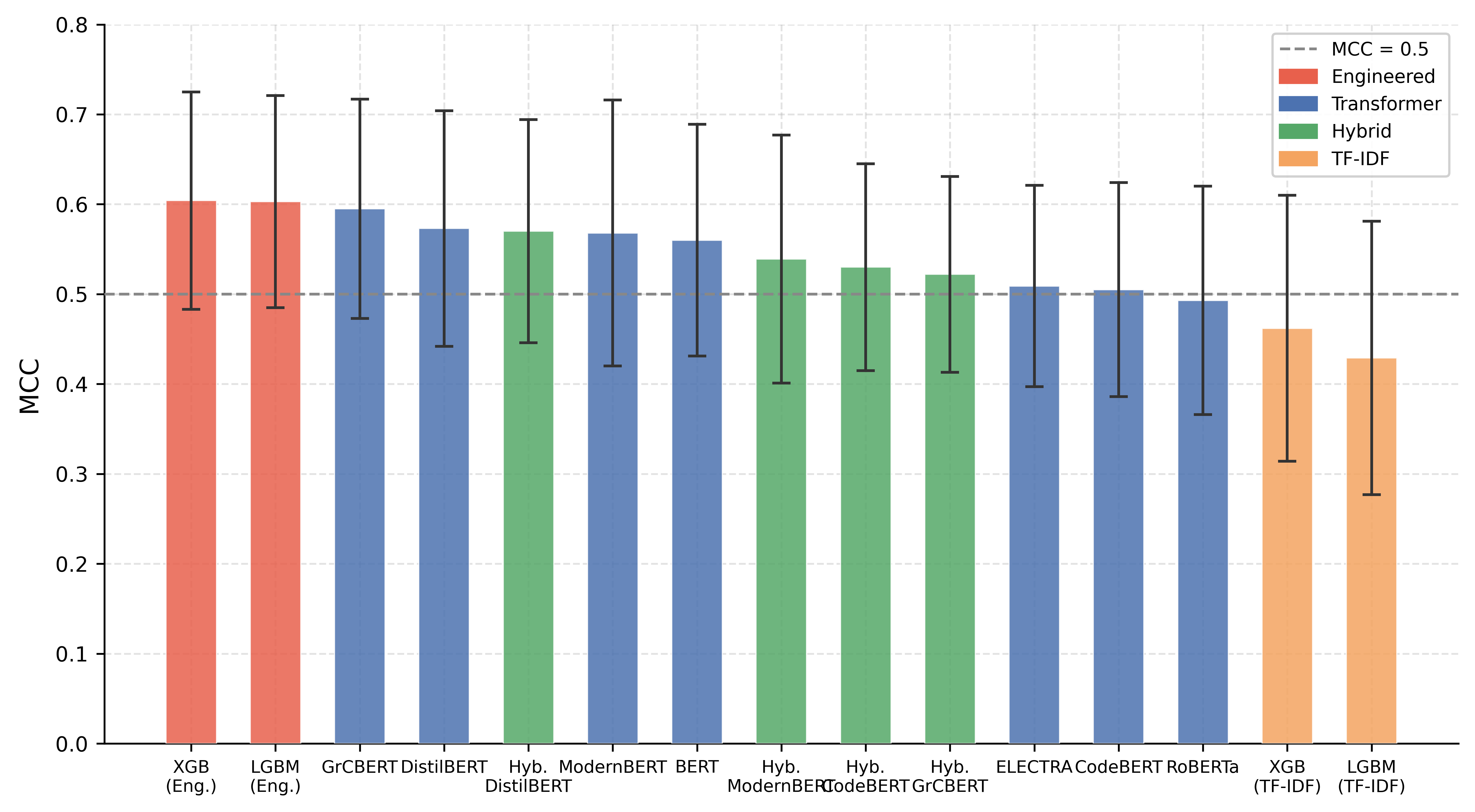}
    \caption{Mean MCC with standard deviation 
    error bars across three folds for all 
    evaluated models. Wider error bars 
    indicate higher variance in performance 
    across different attack type compositions.}
    \label{fig:errorbars}
\end{figure}

\begin{figure}[H]
    \centering
    \includegraphics[width=\columnwidth]{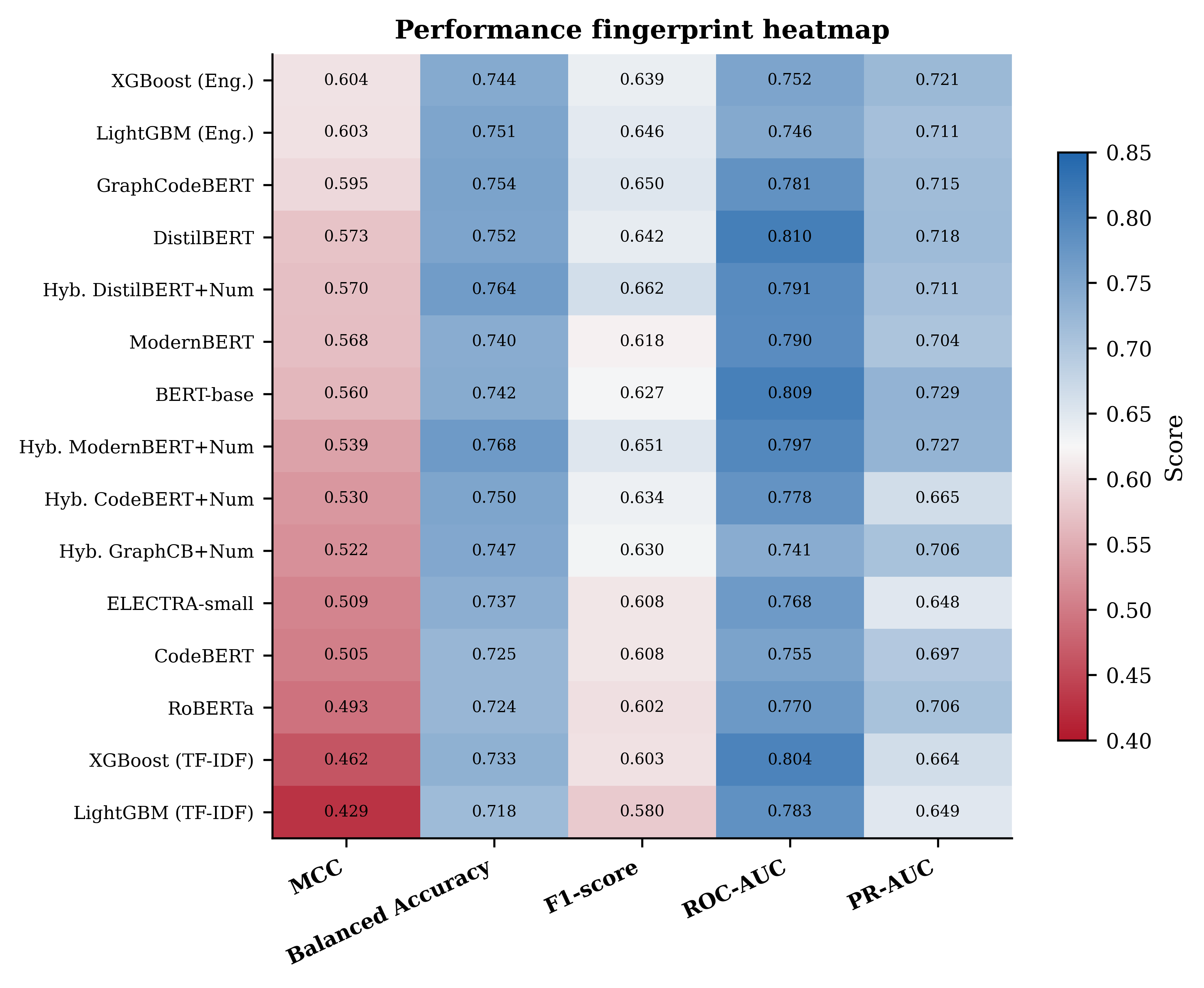}
    \caption{Performance heatmap of all evaluated models across MCC, Balanced Accuracy, F1-score, ROC-AUC, and PR-AUC. The heatmap highlights the strong performance of Transformer models compared with TF-IDF baselines and shows that engineered-feature baselines remain highly competitive.}
    \label{fig:heatmap}
\end{figure}

\begin{table*}[!t]
\centering
\caption{Overall model performance ranked by MCC.}
\label{tab:overall_performance}

\renewcommand{\arraystretch}{1.65}
\setlength{\tabcolsep}{5pt}

\resizebox{\textwidth}{!}{
\begin{tabular}{|p{1.0cm}|p{3.65cm}|p{2.65cm}|p{1.6cm}|p{1.6cm}|p{1.6cm}|p{1.6cm}|p{1.6cm}|}
\hline
\textbf{Rank} & \textbf{Model} & \textbf{Type} & \textbf{MCC} & \textbf{Bal. Acc.} & \textbf{F1} & \textbf{ROC-AUC} & \textbf{PR-AUC} \\
\hline
1 & XGBoost (Eng.) & Baseline & \textbf{0.604} & 0.744 & 0.639 & 0.752 & 0.721 \\
\hline
2 & LightGBM (Eng.) & Baseline & \textbf{0.603} & 0.751 & 0.646 & 0.746 & 0.711 \\
\hline
3 & GraphCodeBERT & Transformer & \textbf{0.595} & 0.754 & 0.650 & 0.781 & 0.715 \\
\hline
4 & DistilBERT & Transformer & 0.573 & 0.752 & 0.642 & 0.810 & 0.718 \\
\hline
5 & Hyb. DistilBERT+Num & Hybrid & 0.570 & 0.764 & 0.662 & 0.791 & 0.711 \\
\hline
6 & ModernBERT & Transformer & 0.568 & 0.740 & 0.618 & 0.790 & 0.704 \\
\hline
7 & BERT-base & Transformer & 0.560 & 0.742 & 0.627 & 0.809 & 0.729 \\
\hline
8 & Hyb. ModernBERT+Num & Hybrid & 0.539 & 0.768 & 0.651 & 0.797 & 0.727 \\
\hline
9 & Hyb. CodeBERT+Num & Hybrid & 0.530 & 0.750 & 0.634 & 0.778 & 0.665 \\
\hline
10 & Hyb. GraphCB+Num & Hybrid & 0.522 & 0.747 & 0.630 & 0.741 & 0.706 \\
\hline
11 & ELECTRA-small & Transformer & 0.509 & 0.737 & 0.608 & 0.768 & 0.648 \\
\hline
12 & CodeBERT & Transformer & 0.505 & 0.725 & 0.608 & 0.755 & 0.697 \\
\hline
13 & RoBERTa & Transformer & 0.493 & 0.724 & 0.602 & 0.770 & 0.706 \\
\hline
14 & XGBoost (TF-IDF) & Baseline-TF & 0.462 & 0.733 & 0.603 & 0.804 & 0.664 \\
\hline
15 & LightGBM (TF-IDF) & Baseline-TF & 0.429 & 0.718 & 0.580 & 0.783 & 0.649 \\
\hline
\end{tabular}
}

\end{table*}

\subsection{Pre-Trained Representations vs. Text Format}
One of the primary objectives of this study was to determine whether performance improvements arise from the structured text representation of the data or from the use of pre-trained Transformer models. To find this, GraphCodeBERT and XGBoost-TF-IDF were evaluated using exactly the same text windows. The results show that GraphCodeBERT achieved an MCC improvement of 0.133 over XGBoost-TF-IDF. Because both models were evaluated using identical inputs, the observed improvement can be attributed to the Transformer's ability to learn contextual relationships and sequential dependencies within packet windows.

\begin{figure}[!t]
    \centering
    \includegraphics[width=\columnwidth]{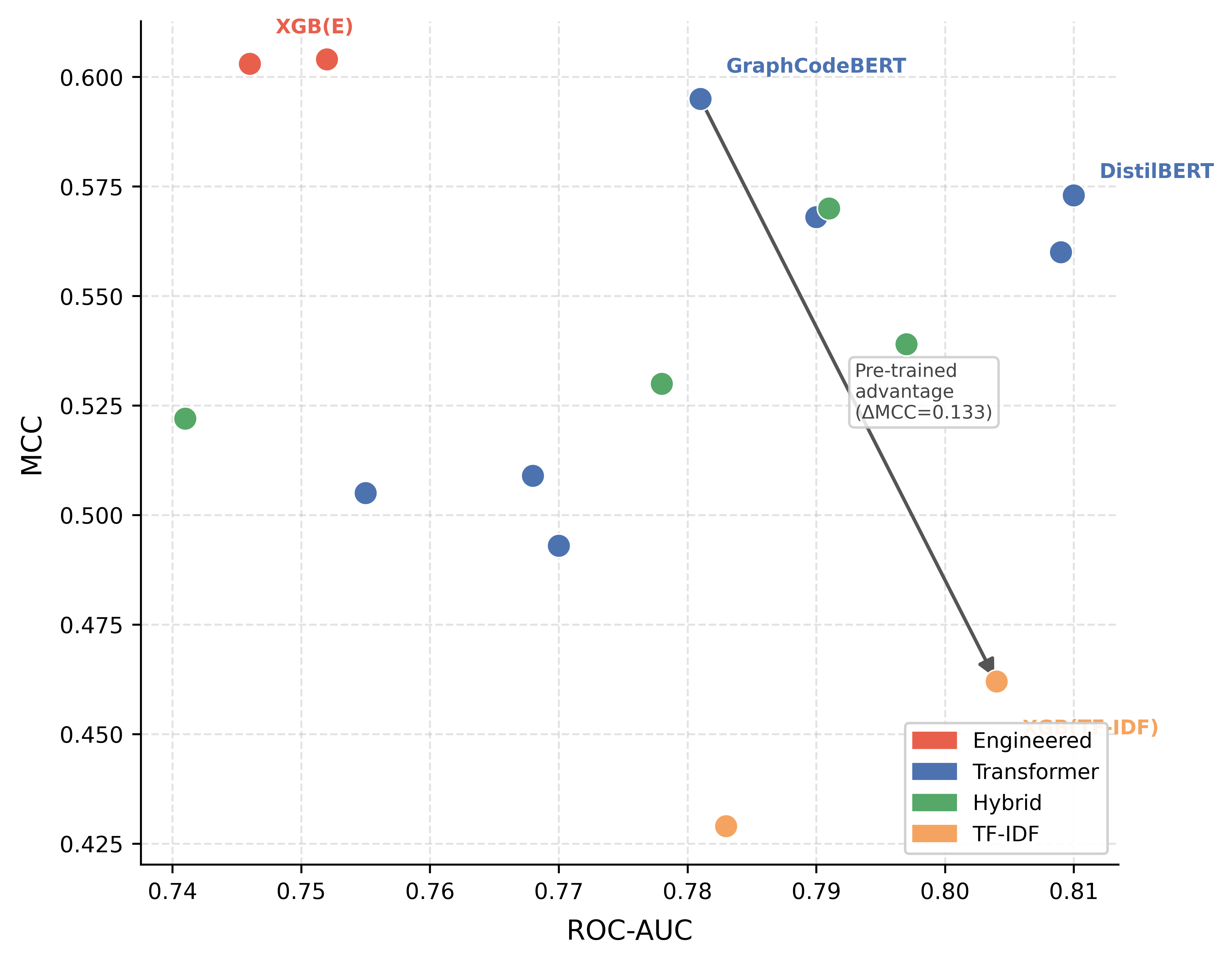}
    \caption{Comparison between MCC and ROC-AUC for the evaluated models. The figure highlights the performance difference between pre-trained Transformer representations and TF-IDF-based text representations using the same structured GOOSE text input.}
    \label{fig:roc_mcc}
\end{figure}

\begin{figure}[H]
    \centering
    \includegraphics[width=\columnwidth]
    {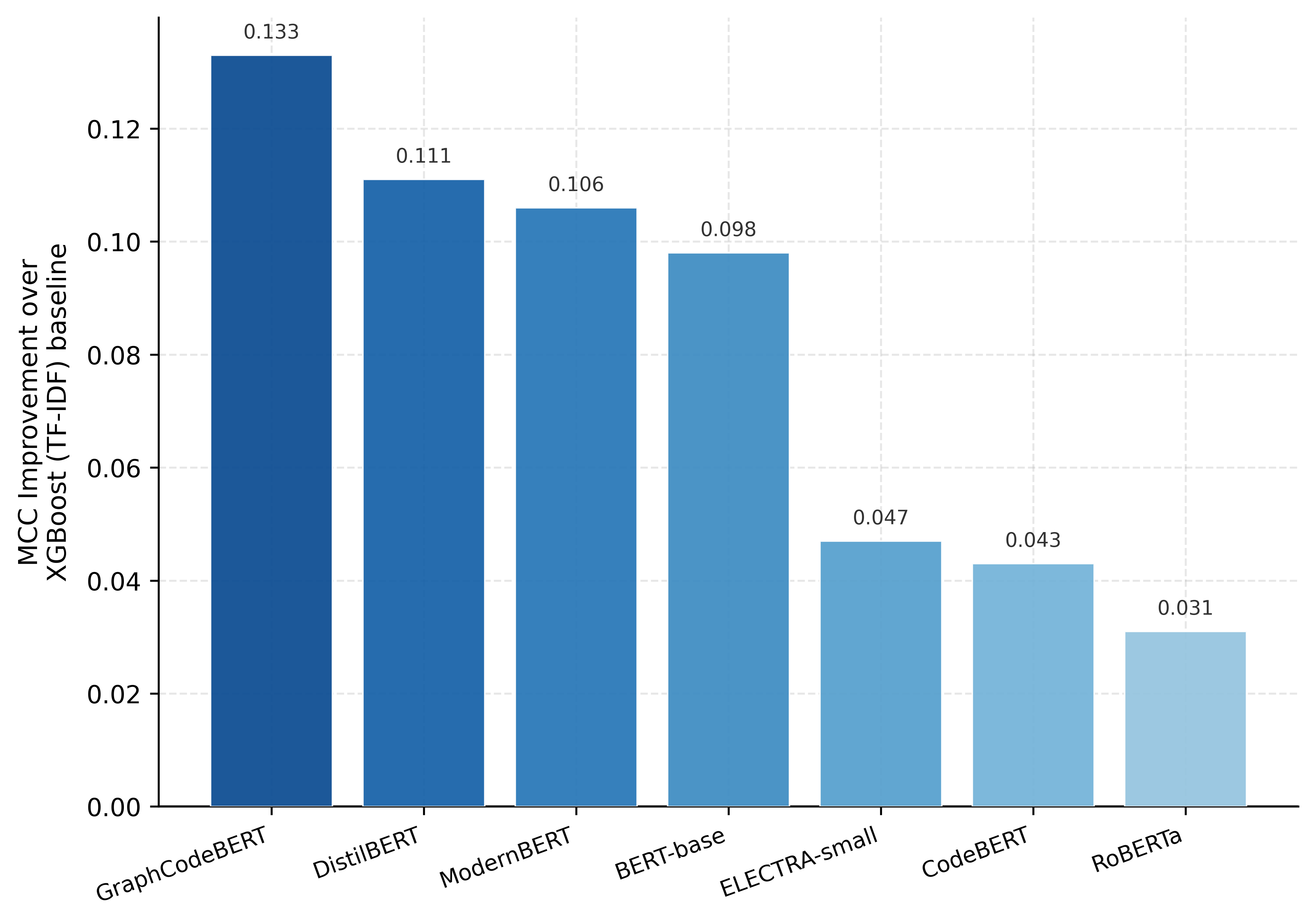}
    \caption{MCC improvement of each fine-tuned Transformer model 
    over the XGBoost (TF-IDF) baseline using identical structured 
    text input $T_w$. GraphCodeBERT achieves the largest gain, 
    confirming that code-aware pre-trained representations provide 
    the strongest contextual advantage over bag-of-words features.}
    \label{fig:mcc_gain}
\end{figure}

\begin{figure}[!t]
    \centering
    \includegraphics[width=\columnwidth]
    {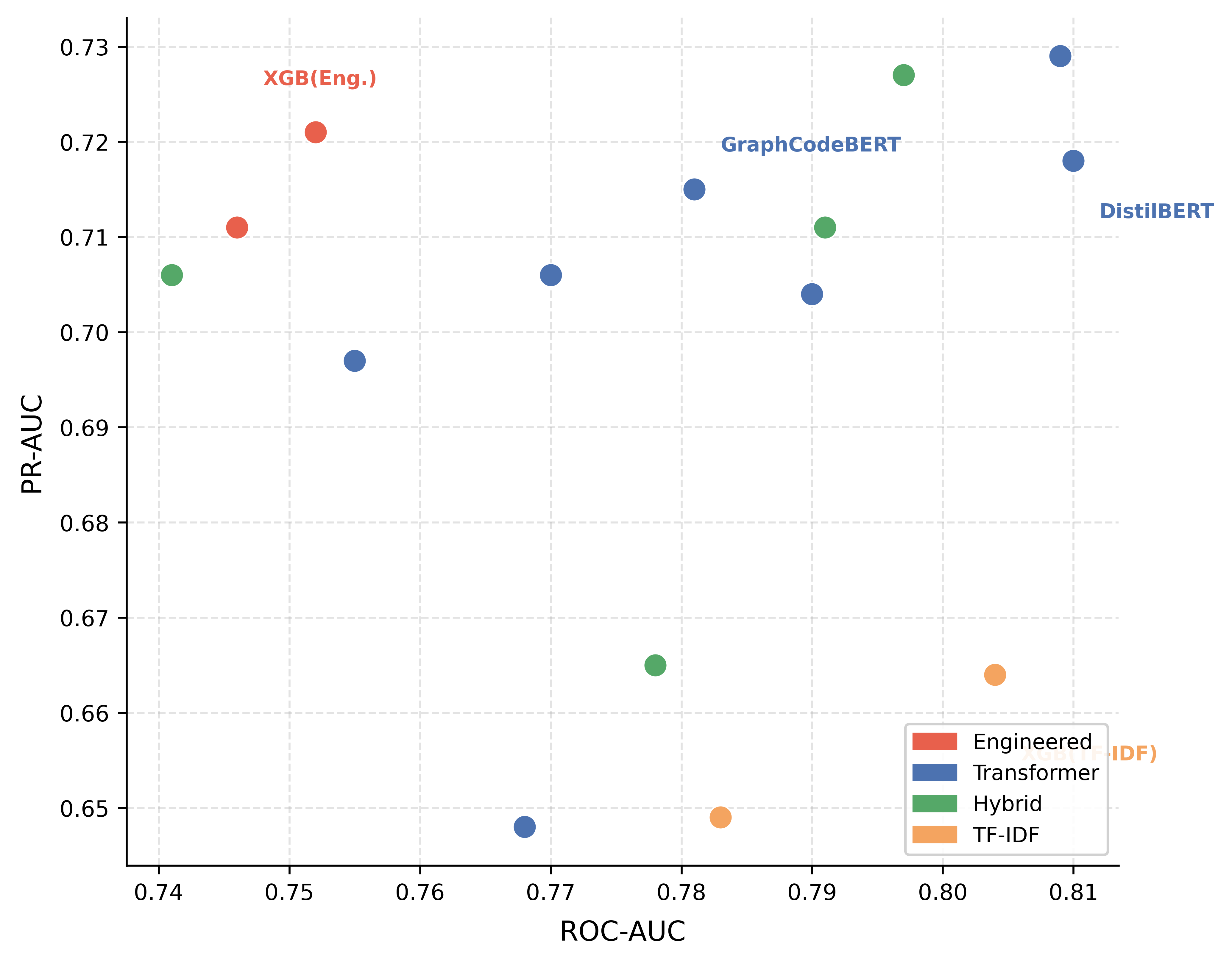}
    \caption{Comparison between PR-AUC and 
    ROC-AUC across model families. PR-AUC 
    captures precision-recall performance 
    under class imbalance and reveals 
    differences not visible in ROC-AUC alone.}
    \label{fig:prauc_rocauc}
\end{figure}

\subsection{Comparison with Expert-Engineered Baselines }
A main question of this study is whether FDIFormer can match the detection performance of expert-engineered feature baselines without relying on any manual protocol feature design. As shown in Table~\ref{tab:overall_performance} and discussed in Section VI-C, GraphCodeBERT achieves statistically equivalent MCC to XGBoost-Engineered. This result is notable because XGBoost-Engineered uses 36 hand-crafted protocol features developed with significant domain expertise, whereas GraphCodeBERT learns directly from the raw structured text representation $T_w$ with no manual feature engineering.

On ROC-AUC, the Transformer models collectively outperform the expert-engineered baselines. DistilBERT achieves the highest ROC-AUC of 0.810 and BERT-base achieves 0.809, compared to XGBoost-Engineered at 0.752 and LightGBM Engineered at 0.746. This gap of approximately 0.058 in ROC-AUC indicates that Transformer models develop a stronger overall ranking capability across detection thresholds, even when their MCC at the optimised threshold is comparable to the engineered baselines.

Among the Transformer models, code-aware architectures (GraphCodeBERT and CodeBERT) consistently outperform general-purpose language models such as RoBERTa and BERT-base on MCC, with GraphCodeBERT ranking third 
overall across all 15 evaluated models. This pattern is consistent with the hypothesis that the key-value structured format of GOOSE packet text representations closely resembles the syntax of structured programming languages 
that GraphCodeBERT and CodeBERT were pre-trained to process \cite{guo2021graphcodebert, feng2020codebert}. The GraphCodeBERT pre-training objective of data flow graph, which learns the relationship between variables and their usage across sequences, might be very relevant to capture sequential dependencies between stNum, sqNum and anomaly flag fields within GOOSE windows.

It is also worth noting that LightGBM-Engineered achieves MCC = 0.603, virtually identical to XGBoost-Engineered, suggesting that the performance of the engineered baseline is primarily determined by the quality of the 36 handcrafted features rather than the choice of gradient boosting algorithm. FDIFormer matches this ceiling without access to any of those features, demonstrating that pre-trained Transformer representations can serve as an effective substitute for domain-specific feature engineering in IEC 61850 GOOSE intrusion detection.

\begin{figure}[H]
    \centering
    \includegraphics[width=\columnwidth]{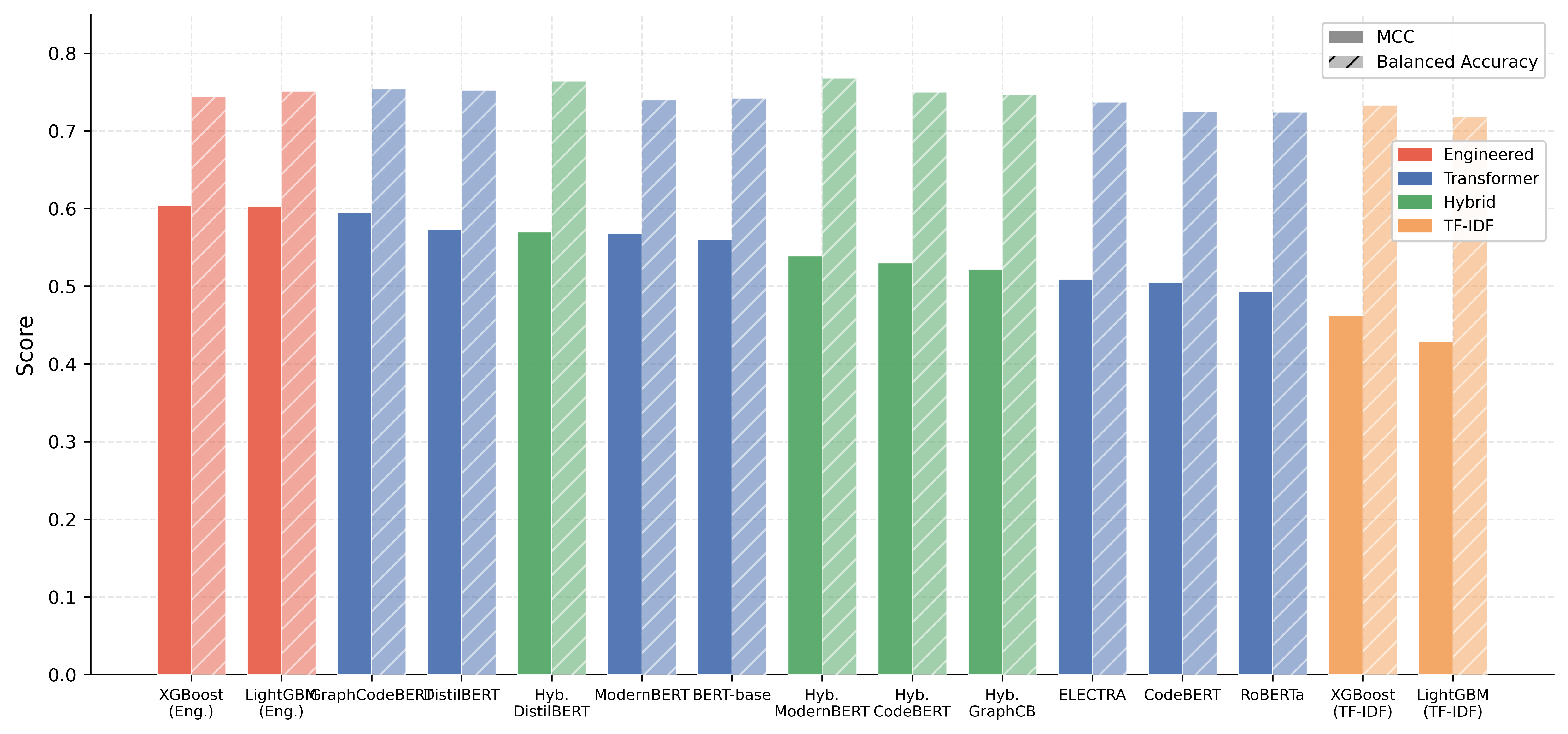}
    \caption{Comparison of MCC and Balanced Accuracy across model families. The figure shows that GraphCodeBERT achieves performance close to expert-engineered XGBoost and LightGBM baselines while avoiding manual protocol feature engineering.}
    \label{fig:mccvsbal}
\end{figure}

Fig.~\ref{fig:mccvsbal} plots MCC against Balanced Accuracy for all 15 models. Figure shows that GraphCodeBERT is in the upper right region together with the engineered baselines, while TF-IDF models are well separated in the lower left area, further validating the advantages of pre-trained representations over bag-of-words text features.

\subsection{Hybrid Model Performance }
The hybrid Transformer-numerical model family was evaluated to determine whether explicit combination of pre-trained Transformer representations with the 36 engineered numerical features $F_w$ could improve detection performance beyond either modality alone. The picture is more complex as shown in Table~\ref{tab:overall_performance}.

Hybrid models show statistically significant improvements on secondary metrics. Hybrid\_DistilBERT+Num outperforms all 15 models (0.662 F1-score) and Hybrid\_ModernBERT+Num outperforms all models (0.768 Balanced Accuracy), with both models outperforming the standalone Transformer models and the engineered-feature baselines on these metrics. These improvements show that the numerical branch is providing additional discriminative signal for some attack scenarios, especially in cases where the Transformer branch alone might produce borderline classification probabilities.

However, on the primary metric MCC, no hybrid model consistently outperforms its corresponding standalone Transformer. Hybrid\_DistilBERT+Num achieves MCC = 0.570, compared to standalone DistilBERT at 0.573. Similarly, 
Hybrid\_GraphCB+Num achieves MCC = 0.522, compared to standalone GraphCodeBERT at 0.595 — a notable decrease. This suggests that the concatenation of Transformer embeddings with numerical features does not provide a consistent MCC benefit and may in some cases introduce noise into the joint representation.

\begin{figure}[H]
    \centering
    \includegraphics[width=\columnwidth]{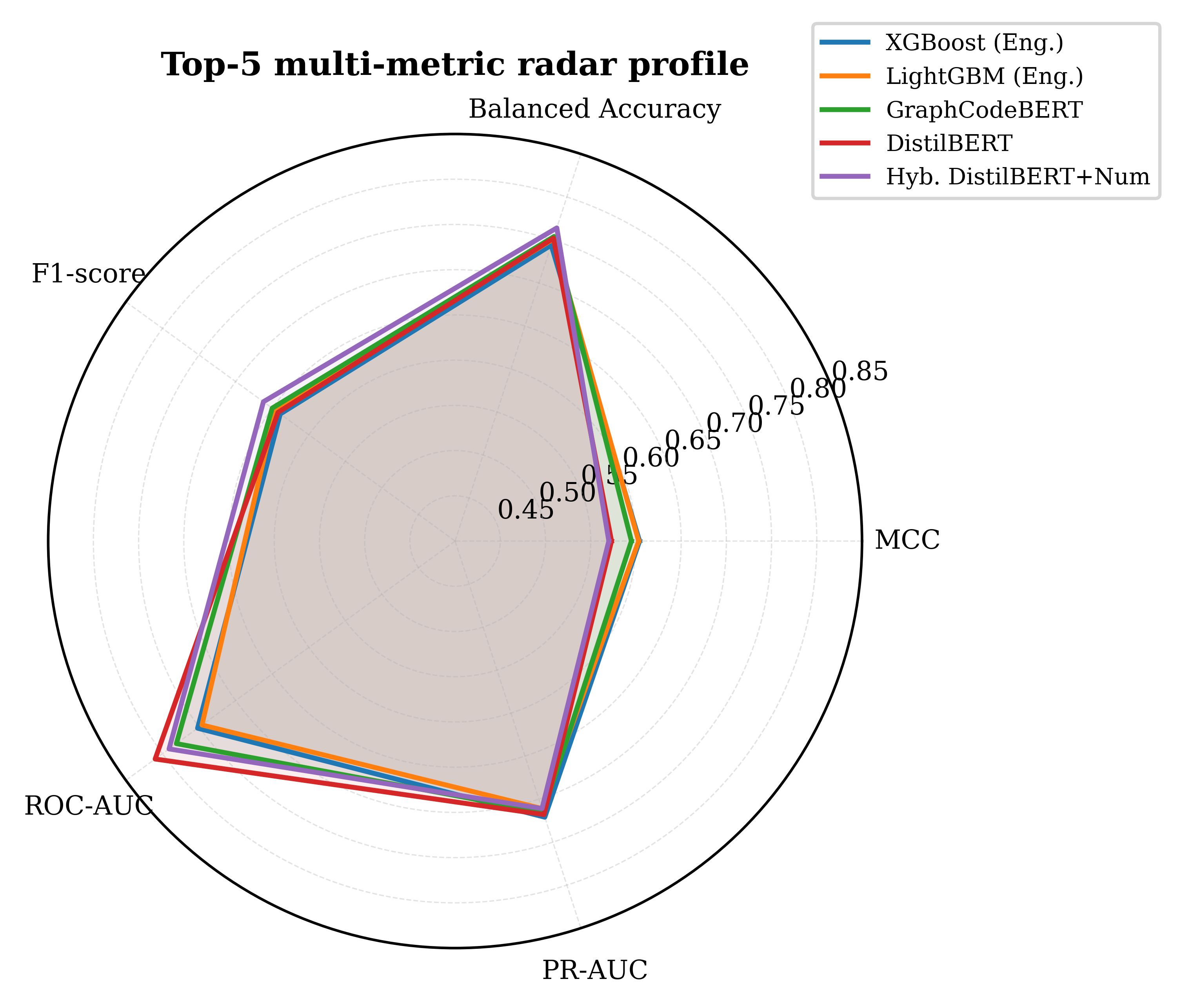}
    \caption{Top-5 multi-metric radar profile comparing the strongest models across MCC, Balanced Accuracy, F1-score, ROC-AUC, and PR-AUC. The figure shows that hybrid models improve some secondary metrics; however, they do not consistently outperform the strongest single Transformer model on MCC.}
    \label{fig:top5_radar}
\end{figure}

A reasonable explanation for such behaviour is that the attack-type prefix encoded in $T_w$ already has some aggregate information based on those same 36 engineered features, including the four binary anomaly flags, length anomaly and payload anomaly indicators. 
This means the Transformer attention mechanism may already be attending to this information via the prefix token, making the explicit numerical branch somewhat redundant. This is in line with the fact that the MCC gap between hybrid and standalone models is largest for GraphCodeBERT — the model that can best extract structural information from the prefix encoding. 
Fig.~\ref{fig:top5_radar} presents the top-5 multi-metric radar profile comparing the strongest models across all five evaluation metrics. The radar plot confirms that hybrid models occupy a different region of the performance space from standalone Transformers — stronger on balanced accuracy and F1-score however not on MCC — and the results indicate that no single model consistently achieves the highest performance across all five evaluation metrics.

\subsection{Attack Type Detectability  }

In Table~\ref{tab:fold_mcc_analysis} we present the MCC breakdown at the fold level for representative models across the three cross-validation folds. The different composition of FDI attack types in the test set for each fold is: Fold 1 includes deletion and addition attack scenarios, Fold 2 includes modification and addition scenarios and Fold 3 includes only modification scenarios. This design provides a systematic evaluation of the model performance across attack types.

The results show a consistent and significant performance degradation from Fold 1 to Fold 3 for all model families. GraphCodeBERT reduces from 0.658 MCC in Fold 1 to 0.455 in Fold 3, a difference of 0.203. Similar drop from 0.690 to 0.465 for XGBoost-Engineered. Even the strongest models fail to reach the MCC = 0.5 threshold in Fold 3, suggesting that modification-only test cases are generally a harder detection task than deletion or addition attacks.

\begin{table}[!t]
\centering
\caption{Fold-Level MCC Analysis by Attack Type Composition}
\label{tab:fold_mcc_analysis}

\footnotesize
\renewcommand{\arraystretch}{1.25}
\setlength{\tabcolsep}{4pt}

\begin{tabular}{|l|c|c|c|c|}
\hline
\textbf{Model} &
\textbf{Fold 1} &
\textbf{Fold 2} &
\textbf{Fold 3} &
\textbf{Mean MCC} \\
&
\textbf{(Del+Add)} &
\textbf{(Mod+Add)} &
\textbf{(Mod only)} &
\\
\hline

GraphCodeBERT & 0.658 & 0.673 & 0.455 & 0.595 \\
\hline

DistilBERT & 0.672 & 0.647 & 0.400 & 0.573 \\
\hline

ModernBERT & 0.685 & 0.624 & 0.396 & 0.568 \\
\hline

BERT-base & 0.664 & 0.609 & 0.408 & 0.560 \\
\hline

XGB (Eng.) & 0.690 & 0.656 & 0.465 & 0.604 \\
\hline

XGB (TF-IDF) & 0.549 & 0.567 & 0.271 & 0.462 \\
\hline

\end{tabular}
\end{table}

\begin{figure}[!t]
    \centering
    \includegraphics[width=\columnwidth]{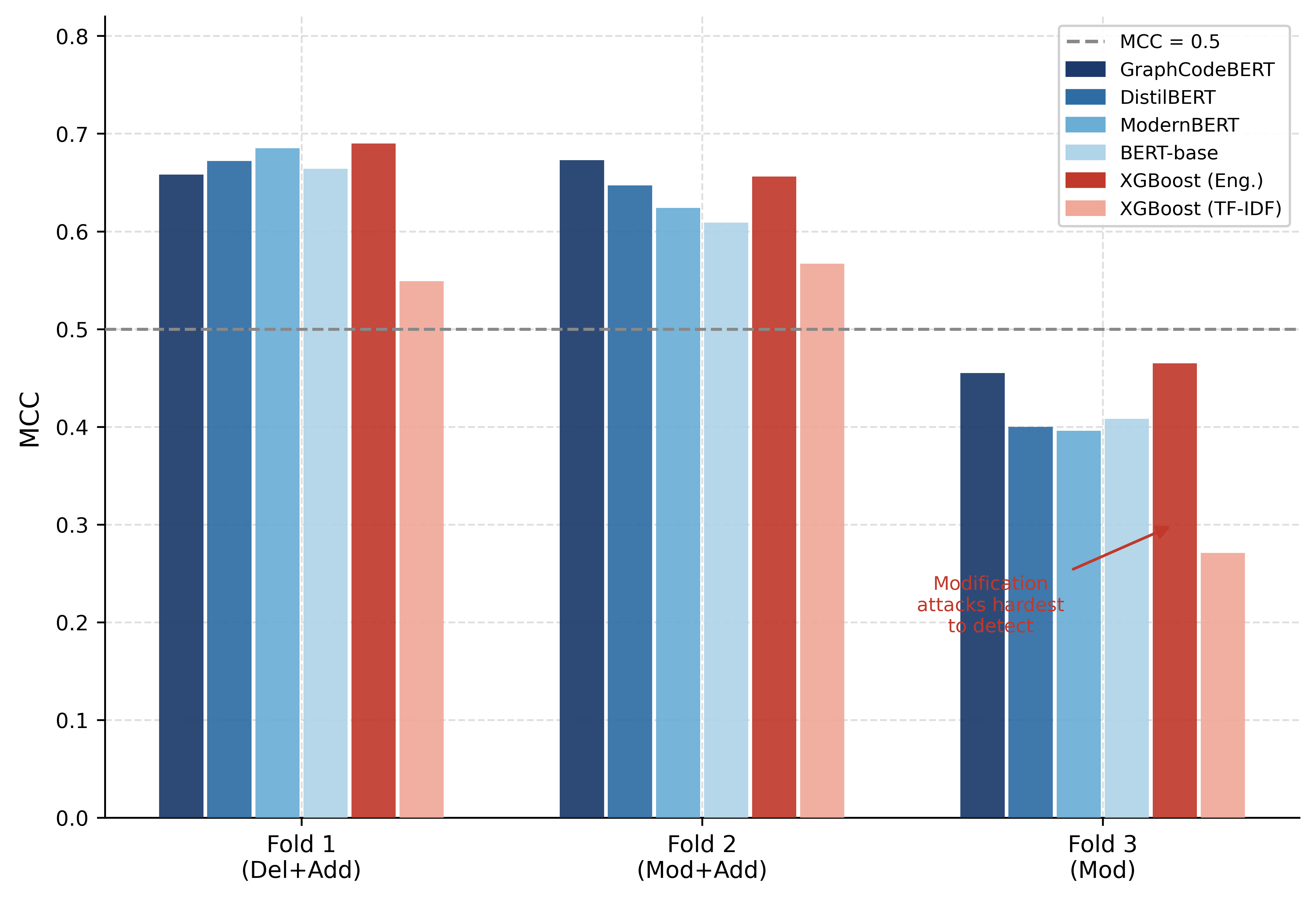}
    \caption{Fold-level MCC comparison across representative model families. The lower performance in Fold 3 indicates that modification-heavy scenarios are more difficult to detect than deletion and addition attack scenarios.}
    \label{fig:foldlevel}
\end{figure}
As discussed, deletion and addition attacks disrupt multiple protocol fields simultaneously, producing strong multi-feature anomaly signals, whereas modification attacks alter only the payload values while preserving the complete structural format of legitimate GOOSE messages. This explains why modification-only scenarios in Fold 3 produce a much weaker and less reliable detection signal across all model families.

Fig.~\ref{fig:foldlevel} visualises the fold-level MCC patterns across model families, clearly showing a performance cliff between Fold 2 and Fold 3 for all evaluated models.

\section{Conclusion}

This paper presented a feature-engineering-free framework for detecting False Data Injection (FDI) attacks in IEC 61850 GOOSE communication using structured textual representations and fine-tuned pre-trained Transformer models. The proposed approach converted GOOSE packet sequences into structured text windows and evaluated multiple Transformer models, classical machine learning baselines, and hybrid architectures under the same experimental framework.

The experimental results showed that GraphCodeBERT achieved the strongest performance among the Transformer models and produced results comparable to the best engineered-feature baselines. In contrast, TF-IDF-based models consistently achieved lower performance, demonstrating that the improvement comes mainly from the pre-trained Transformer representations rather than the text format alone. The results also showed that modification attacks remain the most challenging attack type for all model categories due to their ability to preserve normal packet structure while manipulating payload values.

Overall, the findings indicate that pre-trained Transformer models can provide an effective and generalisable solution for FDI attack detection without relying heavily on manually engineered protocol features. These results should be interpreted with some caution, however, given the limited
number of attack scenarios and cross-validation folds available in the QUT-ZSS-2023-GOOSE dataset and given that inference latency on representative substation hardware was not evaluated in this study. 

Future work will address these limitations by investigating larger and more balanced datasets, explainable AI techniques to support operator trust, real-time edge deployment of compact Transformer models, federated learning across different substations, and extension of the proposed framework to other IEC 61850 communication services and industrial protocols.


%



\ifCLASSOPTIONcaptionsoff
  \newpage
\fi



%

%

\begin{IEEEbiography}[{\includegraphics[width=1in,height=1.25in,clip,keepaspectratio]{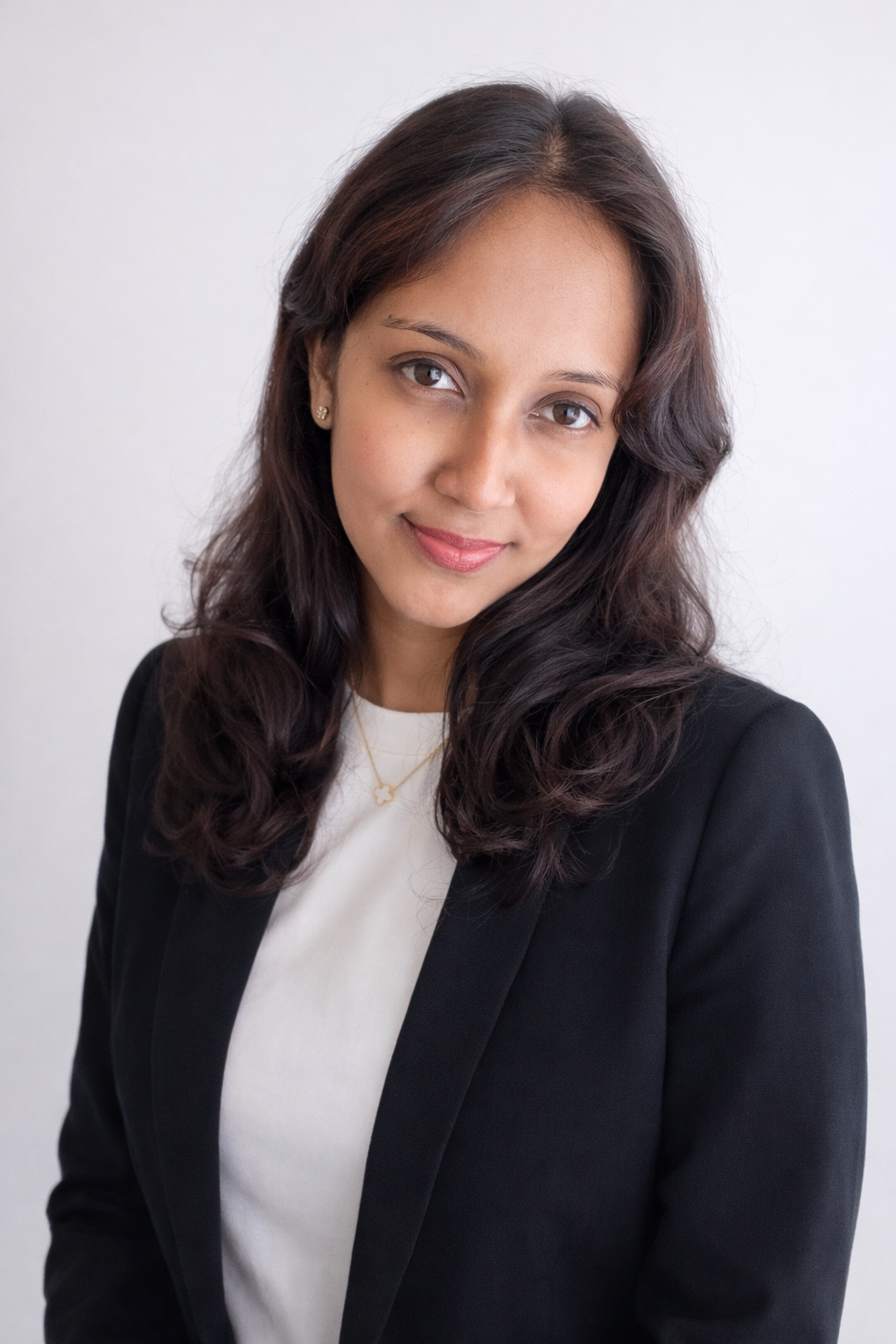}}]{Sandara Sathsarani Wijethunga}
received the B.Sc. (Hons.) degree in Electronics and Telecommunication Engineering from General Sir John Kotelawala Defence University, Sri Lanka, in 2022, graduating with First Class Honours. She is currently pursuing the Master of Applied Artificial Intelligence (Professional) degree at Deakin University, Melbourne, Australia.

Her current research focuses on applying pre-trained Transformer models for cybersecurity in smart grid communication systems, with particular emphasis on False Data Injection (FDI) attack detection in IEC 61850 GOOSE traffic. Her research interests include smart grid cybersecurity, intrusion detection systems, natural language processing, and Transformer-based learning.

\end{IEEEbiography}



\begin{IEEEbiography}[{\includegraphics[width=1in,height=1.25in,clip,keepaspectratio]{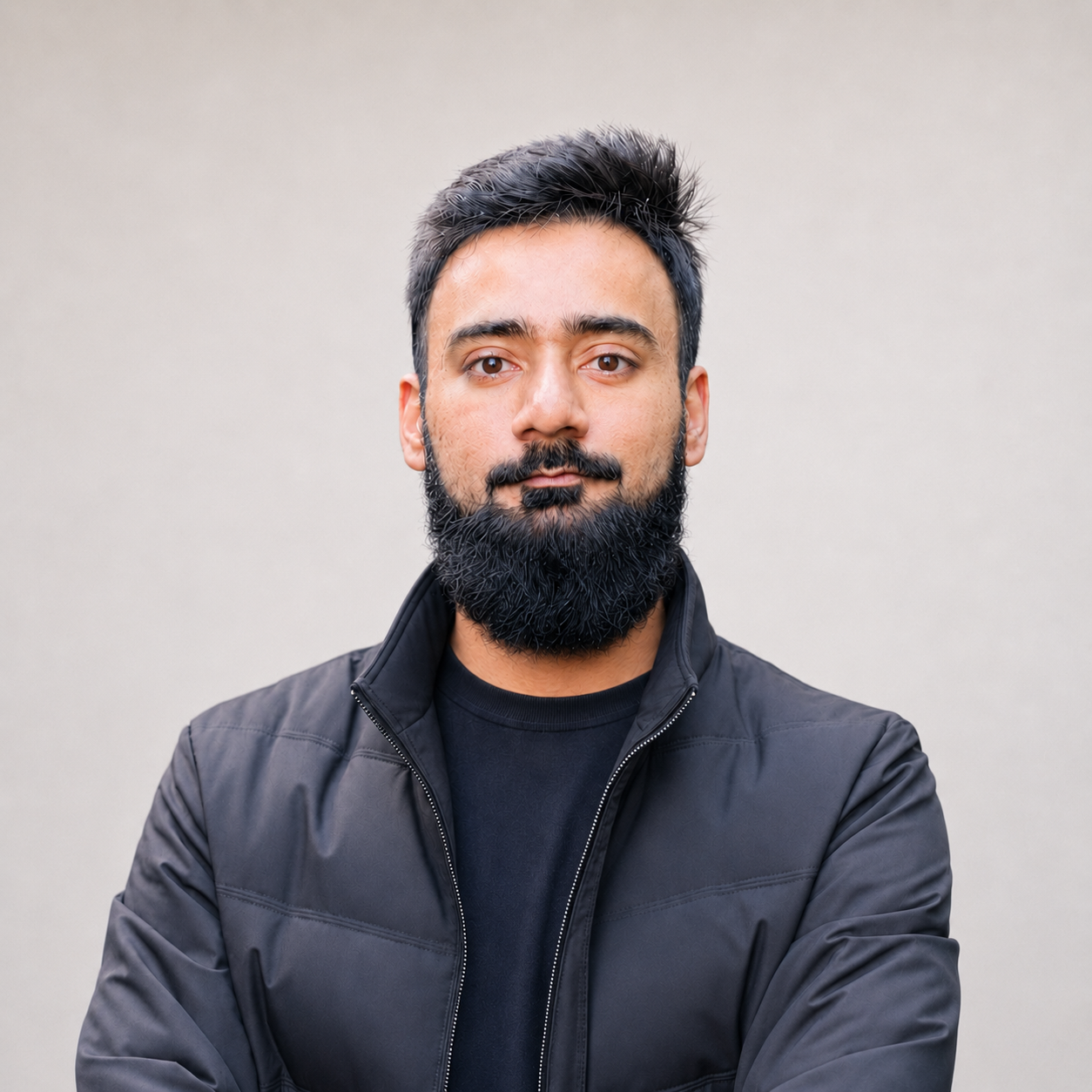}}]{Muneeb Ul Hassan} is currently working as a Lecturer (Assistant Professor) in Cybersecurity at Deakin University, Australia. He previously worked as a Postdoctoral Research Associate in Security and Privacy at Swinburne University of Technology, Australia. He received his Ph.D. from Swinburne University of Technology, Australia, in 2021. He completed his Bachelor of Electrical Engineering at COMSATS Institute of Information Technology, Wah Cantt. He was awarded a Gold Medal by the university for being the top student in the Department of Electrical Engineering. He is also a recipient of the IEEE TCSC 2022 Award for Excellence, Early Career Researcher, in Scalable Computing for his research excellence in privacy preservation for blockchain and decentralized energy systems. In addition, he has won several Top Peer Reviewer Awards from Clarivate Web of Science.
He has published his research in top-tier journals in the field, including IEEE Transactions on Knowledge and Data Engineering, IEEE Transactions on Services Computing, and IEEE Communications Surveys \& Tutorials. His publications have an accumulated impact factor of more than 100. His main research interests include privacy preservation, electric vehicles, blockchain technology, differential privacy, cybersecurity, smart grids, cognitive radio ad hoc networks, artificial intelligence, cloud computing, big data security and privacy, wireless networks, cognitive radio sensor networks, and mobile ad hoc networks.

\end{IEEEbiography}

\begin{IEEEbiography}[{\includegraphics[width=1in,height=1.25in,clip,keepaspectratio]{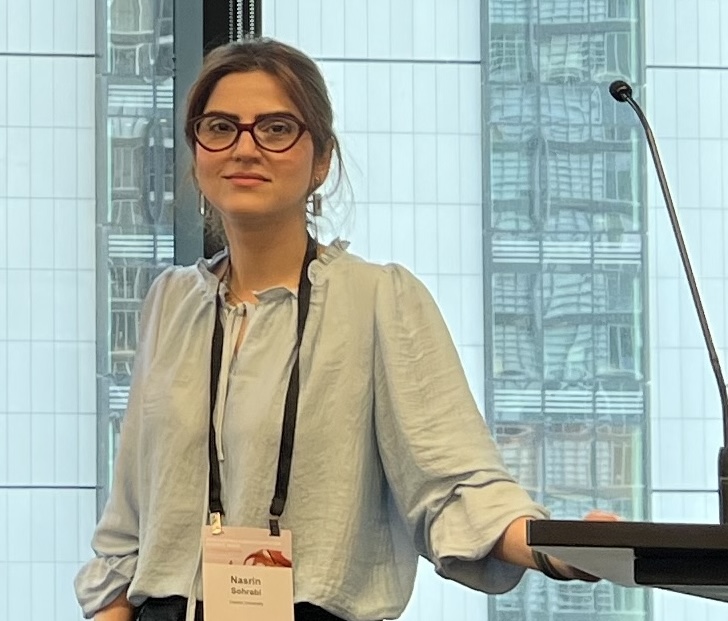}}]{Nasrin Sohrabi} is a Lecturer (assistant Professor) in Cybersecurity at Deakin University, Melbourne. She received her Ph.D. in Computer Science from RMIT University. Her research focuses on systems security, anomaly detection, blockchain forensics, and reliability  of decentralized systems. She has published in leading venues, including IEEE International Conference on Data Engineering (ICDE), IEEE Transactions on Dependable and Secure Computing (TDSC), IEEE Transactions on Services Computing (TSC), IEEE Transactions on Information Forensics and Security (TIFS), ACM Computing Surveys, and Journal of Parallel and Distributed Computing (JPDC). She serves on the program committees of conferences such as ACM International Conference on Information and Knowledge Management (CIKM), IEEE/IFIP International Conference on Dependable Systems and Networks (DSN), FAB, and Australasian Information Security Conference (AISC). In addition, she regularly serves as a reviewer for prestigious journals, including IEEE Transactions on Parallel and Distributed Systems (TPDS), TDSC, TSC, and IEEE Transactions on Knowledge and Data Engineering (TKDE). She is also the founder and leader of the Resilient and Scalable Computing Group (RSCG).

\end{IEEEbiography}




\end{document}